\documentclass[]{mn2e}
\usepackage{epsfig}

\title[M\,33 monitoring. III]{The UK Infrared Telescope M\,33 monitoring
project. III. Feedback from dusty stellar winds in the central square
kiloparsec}
\author[Javadi et al.]{Atefeh Javadi$^{1}$,
                       Jacco Th.\ van Loon$^{2}$,
                       Habib Khosroshahi$^{1}$ and
                       Mohammad Taghi Mirtorabi$^{3}$\\
$^{1}$School of Astronomy, Institute for Research in Fundamental Sciences
      (IPM), P.O.\ Box 19395-5531, Tehran, Iran\\
$^{2}$Astrophysics Group, Lennard-Jones Laboratories, Keele University,
      Staffordshire ST5 5BG, UK\\
$^{3}$Physics Department, Alzahra University, Vanak, 1993891176, Tehran, Iran}
\date{Submitted: December 2012; Resubmitted: March 2013}
\pagerange{\pageref{firstpage}--\pageref{lastpage}}
\pubyear{2013}
\begin{document}
\maketitle
\label{firstpage}
\begin{abstract}
We have conducted a near-infrared monitoring campaign at the UK InfraRed
Telescope (UKIRT), of the Local Group spiral galaxy M\,33 (Triangulum). The
main aim was to identify stars in the very final stage of their evolution, and
for which the luminosity is more directly related to the birth mass than the
more numerous less-evolved giant stars that continue to increase in
luminosity. In this third paper of the series, we measure the dust production
and rates of mass loss by the pulsating Asymptotic Giant Branch (AGB) stars
and red supergiants. To this aim, we combined our time-averaged near-IR
photometry with the multi-epoch mid-IR photometry obtained with the {\it
Spitzer} Space Telescope. The mass-loss rates are seen to increase with
increasing strength of pulsation and with increasing bolometric luminosity.
Low-mass stars lose most of their mass through stellar winds, but even
super-AGB stars and red supergiants lose $\sim40$\% of their mass via a dusty
stellar wind. More than three-quarters of the dust return is oxygenous. We
construct a 2-D map of the mass-return rate, showing a radial decline but
also local enhancements due to agglomerations of massive stars. We estimate a
total mass-loss rate of 0.004--0.005 M$_\odot$ yr$^{-1}$ kpc$^{-2}$, increasing
to $\sim0.006$ M$_\odot$ yr$^{-1}$ kpc$^{-2}$ when accounting for eruptive mass
loss (e.g., supernov{\ae}); comparing this to the current star formation rate
of $\sim0.03$ M$_\odot$ yr$^{-1}$ kpc$^{-2}$ we conclude that star formation in
the central region of M\,33 can only be sustained if gas is accreted from
further out in the disc or from circum-galactic regions.
\end{abstract}

%=========================================================================== 1
\section{Introduction}

Galactic evolution is driven at the end-points of stellar evolution, where
copious mass loss returns chemically-enriched and sometimes dusty matter back
to the interstellar medium (ISM); the stellar winds of evolved stars and the
violent deaths of the most massive stars also inject energy and momentum into
the ISM, generating turbulence and galactic fountains when superbubbles pop as
they reach the ``surface'' of the galactic disc. The evolved stars are also
excellent tracers, not just of the feedback processes, but also of the
underlying populations, that were formed from millions to billions of years
prior to their appearance. The evolved phases of evolution generally represent
the most luminous, and often the coolest, making evolved stars brilliant
beacons at IR wavelengths, where it is also easier to see them deep inside
galaxies as dust is more tranpsarent at those longer wavelengths than in the
optical and ultraviolet where their main-sequence progenitors shine. The final
stages of stellar evolution of stars with main-sequence masses up to $M\sim30$
M$_\odot$ -- Asymptotic Giant Branch (AGB) stars and red supergiants -- are
characterised by strong radial pulsations of the cool atmospheric layers,
rendering them identifiable as long-period variables (LPVs) in photometric
monitoring campaigns spanning months to years (e.g., Whitelock, Feast \&
Catchpole 1991; Wood 2000).

The Local Group galaxy Triangulum (Hodierna 1654) -- hereafter referred to as
M\,33 (Messier 1771) -- offers us a unique opportunity to study stellar
populations, their history and their feedback across an entire spiral galaxy
and in particular in its central regions, that in our own Milky Way are
heavily obscured by the intervening dusty Disc (van Loon et al.\ 2003;
Benjamin et al.\ 2005). Our viewing angle with respect to the M\,33 disc is
more favourable (56--$57^\circ$ -- Zaritsky, Elston \& Hill 1989; Deul \& van
der Hulst 1987) than that of the larger M\,31 (Andromeda), whilst the distance
to M\,33 is not much different from that to M\,31 ($\mu=24.9$ mag -- Bonanos
et al.\ 2006). Large populations of AGB stars have been identified in M\,33
(Cioni et al.\ 2008), and red supergiants up to progenitor masses in excess of
20 M$_\odot$ (Drout, Massey \& Meynet 2012). Many of them are dusty LPVs
(McQuinn et al.\ 2007; Thompson et al.\ 2009), and these have been found also
in the central parts of M\,33 (Javadi, van Loon and Mirtorabi 2011a).

The main objectives of our project are described in Javadi, van Loon \&
Mirtorabi (2011c): to construct the mass function of LPVs and derive from this
the star formation history in M\,33; to correlate spatial distributions of the
LPVs of different mass with galactic structures (spheroid, disc and spiral arm
components); to measure the rate at which dust is produced and fed into the
ISM; to establish correlations between the dust production rate, luminosity,
and amplitude of an LPV; and to compare the {\it in situ} dust replenishment
with the amount of pre-existing dust. Paper I in the series presented the
photometric catalogue of stars in the inner square kpc (Javadi et al.\ 2011a),
and Paper II presented the galactic structure and star formation history in
the inner square kpc (Javadi, van Loon and Mirtorabi 2011b). This is Paper
III, describing the mass-loss mechanism and dust production rate in the inner
square kpc. Subsequent papers in the series will cover the extension to a
nearly square degree area covering much of the M\,33 optical disc.

%=========================================================================== 2
\section{The data and methods}

To derive the mass-loss rates of the red giant variables we follow a two-step
approach. First, we model the spectral energy distributions (SEDs) of near-IR
variables for which mid-IR counterparts have been identified. We use these
results to construct relations between the dust optical depth and bolometric
corrections on the one hand, and near-IR colours on the other. Then, we apply
those relations to the red giant variables for which no mid-IR counterpart was
identified, to derive their mass-loss rates too.

%------------------------------------------------------------------------- 2.1
\subsection{The near-IR data}

Images in the J, H and K$_{\rm s}$ bands were obtained with the United Kingdom
InfraRed Telescope (UKIRT) with the UIST instrument, covering an area of
approximately $4^\prime\times4^\prime$ corresponding to a square kpc at the
distance of M\,33. Observations in the K$_{\rm s}$ band were made on about a
dozen occasions over the period of 2003--2007. The data, point spread function
(PSF) fitting photometry with {\sc DAOphot} (Stetson 1987) and variability
analysis are described in Paper I. The full, publicly available catalogue
contains 18\,398 stars; 812 of these were identified as variables on the basis
of the near-IR photometry alone, most of them pulsating red giant stars (AGB
stars and red supergiants).

%------------------------------------------------------------------------- 2.2
\subsection{The mid-IR data}

Images in the 3.6, 4.5, 5.8 and 8 $\mu$m bands were obtained with the {\it
Spitzer} Space Telescope with the IRAC instrument, on six occasions over the
period of 2004--2006. We use the photometric catalogue of McQuinn et al.\
(2007), which is based on PSF fitting with {\sc DAOphot} to the first five
epochs and excludes the less sensitive 5.8 $\mu$m band. Thompson et al.\
(2009) have published a somewhat deeper catalogue in the 3.6 and 4.5 $\mu$m
bands including the sixth epoch, but they have not made publicly available
their photometry obtained at longer wavelengths and we therefore prefer
working with the homogeneous catalogue of McQuinn et al. We cross-matched our
near-IR catalogue to their catalogue (see Paper I for details) and found 523
matches in the 3.6 $\mu$m band, 471 at 4.5 $\mu$m and 107 at 8 $\mu$m (Fig.\
1).

%
% FIGURE 1
%
\begin{figure}
\centerline{\psfig{figure=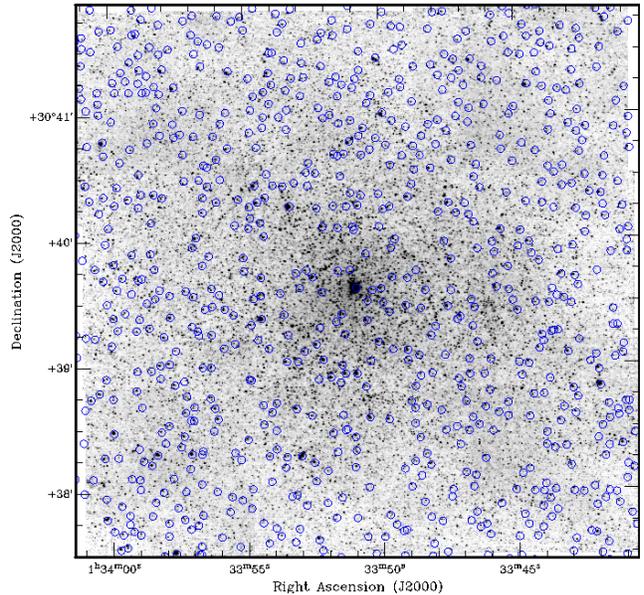,width=84mm}}
\caption[]{K$_{\rm s}$ band image of the central square kpc of M\,33, overlain
with the mid-IR sources from McQuinn et al.\ (2007).}
\end{figure}

We noticed (in Paper I) a discrepancy between the [3.6]--[4.5] colours and
theoretical isochrones (from Marigo et al.\ 2008), with the faintest stars
being abnormally blue. This is not apparent in the colour--magnitude diagram
of M\,33 as a whole (see McQuinn et al.\ 2007) and is most likely related to
the specific challenges encountered in the crowded central regions of the
galaxy (though strong absorption in the fundamental band of carbon monoxide
can depress the 4.5 $\mu$m brightness, which is a challenge for models to
predict accurately). The PSF of the mid-IR data has a full width at half
maximum of $1\rlap{.}^{\prime\prime}7$, which is more than twice as large as
that of the near-IR data.

To gain an appreciation of the severity and effects of blending, we performed
a simple simulation. First, we picked an individual K$_{\rm s}$ band frame and
removed all detected stars from it with {\sc DAOphot/allstar} (Stetson 1987).
Next, we placed each of these stars back into the frame, at their original
positions but using a $1\rlap{.}^{\prime\prime}7$ PSF to mimic the appearance
at 3.6 $\mu$m. We assumed that none of these stars have an infrared excess due
to circumstellar dust, so their 3.6 $\mu$m counts can be derived from their
K$_{\rm s}$ magnitude by applying a typical K$_{\rm s}$--[3.6] colour of a
few tenths of a magnitude (see figure 4 in van Loon, Marshall \& Zijlstra
2005). Then, we used {\sc DAOphot/allstar} and {\sc DAOmaster} to detect and
photomeasure the stars on this pseudo-3.6 $\mu$m frame.

%
% FIGURE 2
%
\begin{figure}
\centerline{\psfig{figure=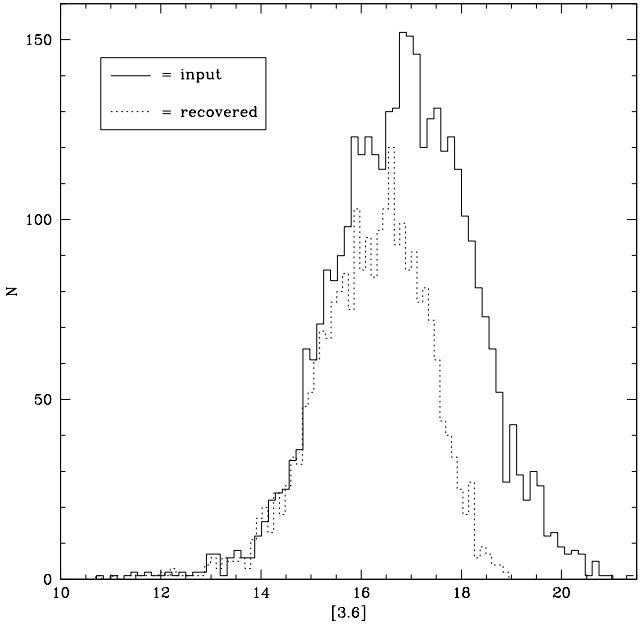,width=84mm}}
\caption[]{Distributions of input magnitudes (solid histogram) and recovered
magnitudes (dotted), derived from a pseudo-3.6 $\mu$m frame simulated on the
basis of a real K$_{\rm s}$ band frame.}
\end{figure}

The distributions of input and recovered pseudo-3.6 $\mu$m magnitudes (Fig.\
2) demonstrate that in the crowded central regions of M\,33 the 3.6 $\mu$m
photometry becomes severely incomplete for stars fainter than about magnitude
16, well before the K$_{\rm s}$ band data. Hence our simulation also accounts
for unresolved emission at 3.6 $\mu$m from stars too faint to be detected
individually. The recovery rate of stars brighter than magnitude 16 is very
high.

%
% FIGURE 3
%
\begin{figure}
\centerline{\psfig{figure=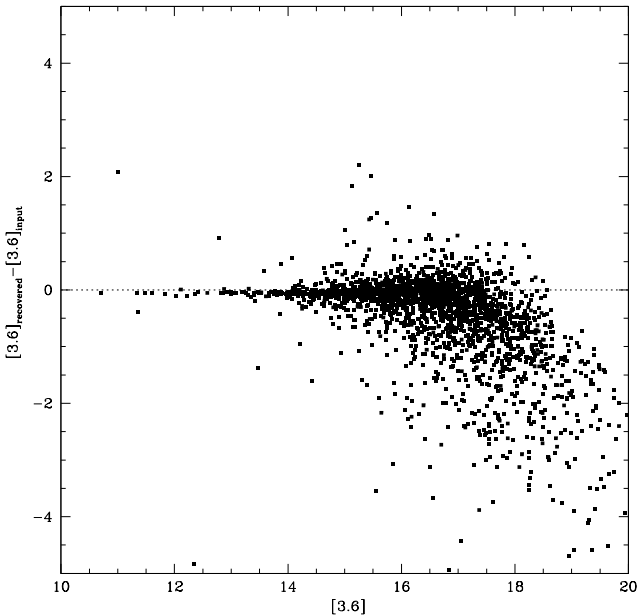,width=84mm}}
\caption[]{Difference between input magnitude and recovered magnitude vs.\
input magnitude, derived from a pseudo-3.6 $\mu$m frame simulated on the basis
of a real K$_{\rm s}$ band frame.}
\end{figure}

Stars brighter than magnitude 16 are recovered with reliable photometry, at an
accuracy (much) better than 0.3 mag and showing no systematic offset (Fig.\
3). Visual inspection of the few bright sources with large photometric
discrepancies revealed that most of these stars were located near the very
nucleus of M\,33 or near the edge of the frame. Stars fainter than magnitude
16 start showing the effects of blending, leading to a reduction in the number
of stars that are recovered (Fig.\ 2) and a systematic over-estimation of
their brightness (Fig.\ 3).

%
% FIGURE 4
%
\begin{figure}
\centerline{\psfig{figure=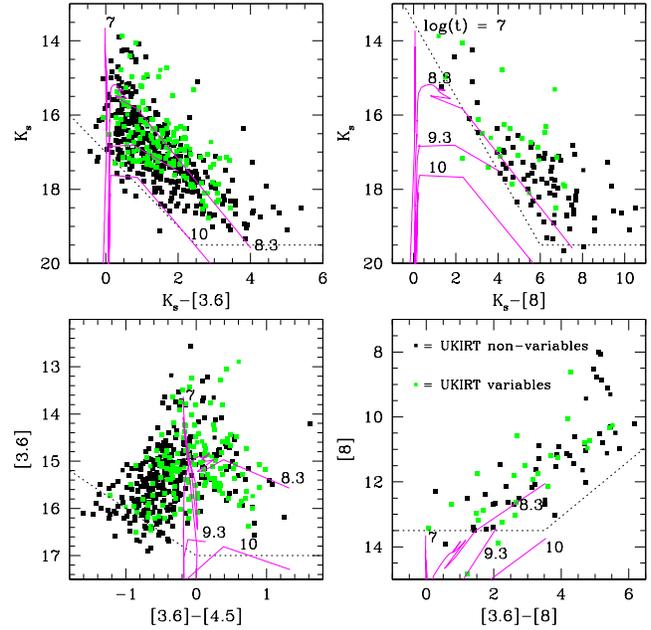,width=84mm}}
\caption[]{Near- and mid-IR colour--magnitude diagrams, with approximate
detection boundaries (dotted lines) and overlain with theoretical isochrones
from Marigo et al.\ (2008) for ages of $\log t=7$ ($t=10$ Myr), 8.3 (200 Myr),
9.3 (2 Gyr) and 10 (10 Gyr). UKIRT variable stars are plotted in green.}
\end{figure}

A useful number of stars (107) are detected at 8 $\mu$m because of excess
emission from circumstellar dust, which allows derivation of their mass-loss
rates. The stars form a branch of increasing mid-IR brightness with increasing
near-/mid-IR colour (Fig.\ 4). The brightest mid-IR objects, $[8]<10$ mag,
with very red colours, ($[3.6]-[8])\approx$ 5 mag, deviate from this main
branch. These have not been identified by us as variable stars, and could be
non-stellar in origin, for instance background galaxies or compact H\,{\sc ii}
regions within M\,33. We will come back to these and other deviant sources at
the end of Section 2.3 and in Section 3.1.

%------------------------------------------------------------------------- 2.3
\subsection{Modelling the spectral energy distributions}

%
% FIGURE 5
%
\begin{figure}
\centerline{\psfig{figure=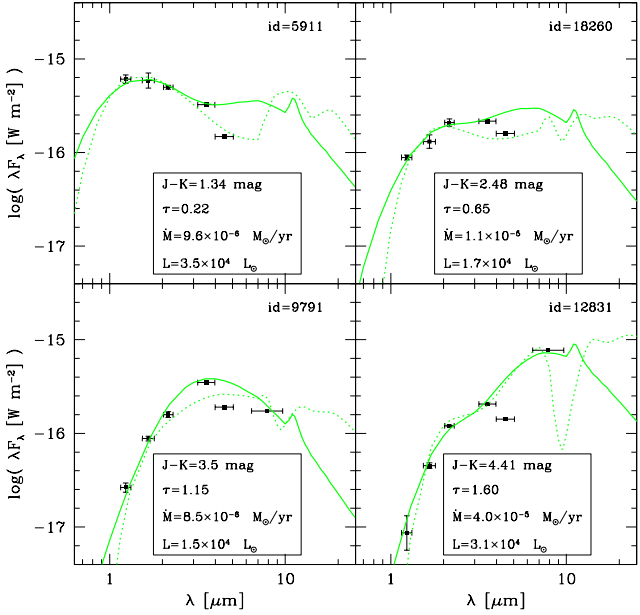,width=84mm}}
\caption[]{Near- and mid-IR photometry of examples of carbon stars in the
centre of M\,33, affected by various levels of mass loss. The horizontal
``errorbars'' on the data represent the width of the photometric passbands.
The solid lines are the best matching SEDs modelled with {\sc dusty}. The
dotted lines are best matching fits using silicates, for comparison.}
\end{figure}

%
% FIGURE 6
%
\begin{figure}
\centerline{\psfig{figure=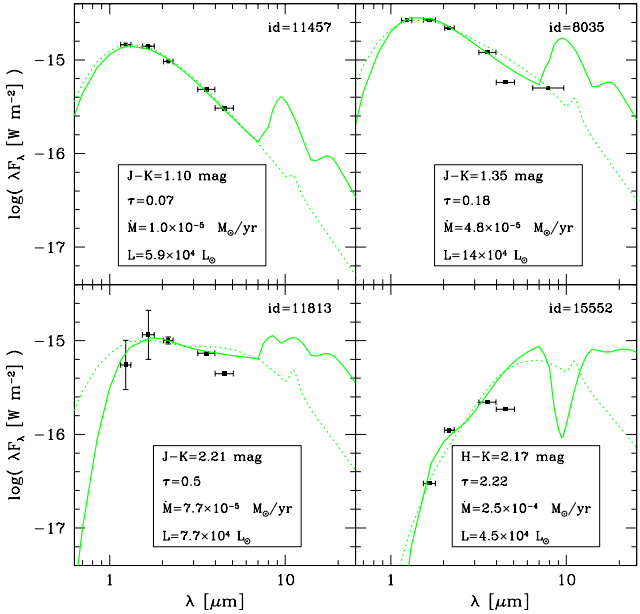,width=84mm}}
\caption[]{As Figure 5 but for examples of M-type stars. The dotted lines are
best matching fits using amorphous carbon and silicon carbide, for
comparison.}
\end{figure}

%
% TABLE 1
%
\begin{table}
\caption{UKIRT ID No.\ (Paper I), optical depth ($\tau$), luminosity ($L$) and
mass-los rate ($\dot{M}$) determined by modelling of the SED with {\sc dusty}.
Note that there is a great deal of uncertainty in the classification into
carbon and M-type stars. (The full table is available electronically.)}
\begin{tabular}{rccc}
\hline\hline
ID   & $\tau$ & $\log L$ (L$_\odot$) & $\log\dot{M}$ (M$_\odot$ yr$^{-1}$) \\
\hline
\multicolumn{4}{l}{\it carbon stars}                                       \\
 629 & 0.07   & 4.45                 & $-5.52$                             \\
1003 & 0.03   & 4.42                 & $-6.43$                             \\
1590 & 0.06   & 4.42                 & $-5.95$                             \\
...  & ...    & ...                  & ...                                 \\
\multicolumn{4}{l}{\it M-type stars}                                       \\
  97 & 0.30   & 3.85                 & $-5.25$                             \\
 901 & 0.20   & 4.15                 & $-4.93$                             \\
2323 & 0.10   & 3.92                 & $-5.62$                             \\
...  & ...    & ...                  & ...                                 \\
\hline
\end{tabular}
\end{table}

We modelled the SEDs of all UKIRT variable stars with measurements in at least
two near-IR bands (K$_{\rm s}$ and J and/or H) and two mid-IR bands (3.6, 4.5
and/or 8 $\mu$m), with the publicly available dust radiative transfer code
{\sc dusty} (based on Ivezi\'c \& Elitzur 1997). Because of the scarcity of
the photometry we fixed the input temperatures of the star and of the dust at
the inner edge of the circumstellar envelope, at 3000 and 900 K, respectively.
The density structure is assumed to follow from the analytical approximation
for radiatively driven winds (Ivezi\'c \& Elitzur 1997). Based on the
estimated birth mass of the star (see Paper II) we used amorphous carbon dust
(Hanner 1988) and a small amount of silicon carbide (P\'egouri\'e 1988) for
$1.5<M/{\rm M}_\odot<4$, and astronomical silicates (Draine \& Lee 1984) for
all other mass ranges, with a gas-to-dust mass ratio of $\psi=200$. The
optical depth ($\tau$) was varied and the luminosity ($L$) was scaled until an
acceptable match was obtained, which was decided by visual inspection. The
optical depth and luminosity, and hence mass-loss rate ($\dot{M}$), were
determined for 58 carbon stars and 35 M-type stars (Table 1); examples are
presented in Figures 5 (carbon stars) and 6 (M-type stars).

Given the limited number of free parameters even a small number of data points
constrain the model fit quite well. In all panels of Figs.\ 5 and 6 we not
only show the best fit for the preferred dust species but also that for the
alternative dust species. While it is not always possible to tell, on the
basis of the fit, which species of dust is present, the preferred dust species
often yields better fits and hardly ever yields worse fits. Our study spans
several orders of magnitude in both mass-loss rate and luminosity, and the
uncertainties resulting from the assumptions in the fitting procedures are
thus relatively unimportant. A heavily dust-enshrouded pulsating giant star
will still be fitted with a model with high optical depth regardless whether
it employs carbonaceous or oxygenous dust. We do note, though, that often the
4.5-$\mu$m datum is anomalously faint compared to adjacent bands. This may be
due to molecular absorption, which was not included in our SED modelling. The
fundamental ro-vibrational band of CO at 4.6 $\mu$m can be strong especially
in the extended atmospheres of pulsating red giant stars (cf.\ Nowotny et al.\
2013). The 3-$\mu$m C$_2$H$_2$+HCN band is very strong in carbon stars, but it
falls largely outside the IRAC 3.6-$\mu$m bandpass; the 3.8-$\mu$m C$_2$H$_2$
band, while strong in metal-poor carbon stars (van Loon et al.\ 2006, 2008),
is not expected to be nearly as strong in the carbon stars in the central
regions of M\,33 that are of near-solar metallicity.

While this approach may seem crude (by necessity), we shall see that the
results compare favourably with those obtained for similar populations of
stars in the Magellanic Clouds.

%
% FIGURE 7
%
\begin{figure}
\centerline{\psfig{figure=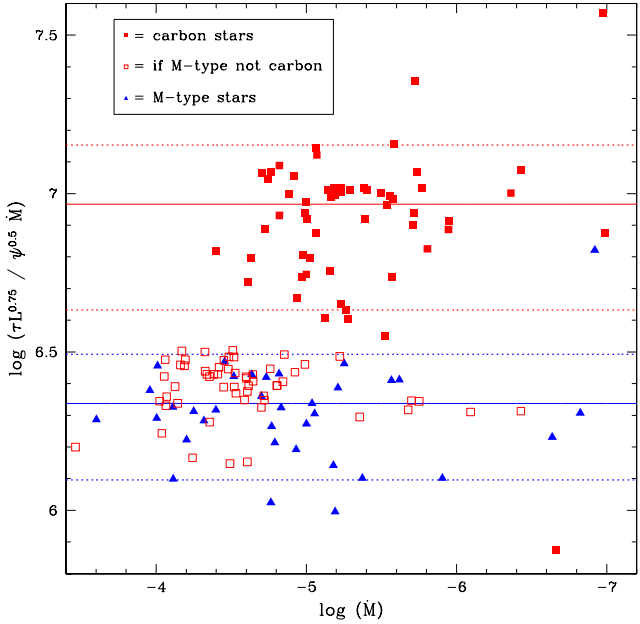,width=84mm}}
\caption[]{Combinations of the optical depth ($\tau$), luminosity ($L$),
gas-to-dust mass ratio ($\psi$) and mass-loss rate ($\dot{M}$) obey scaling
relations derived from the density structure of radiation-driven dust winds
such that the plotted combination is constant to well within a factor two; the
horizontal lines mark the median values (solid) and $\pm$ standard deviation
(pairs of dotted lines). The open red squares show the results if the carbon
stars are presumed to be oxygen-rich instead. The difference between carbon
and M-type stars is due to the different optical properties of the dust
grains.}
\end{figure}

The self-similarity of radiation-driven winds leads to scaling relations
(Ivezi\'c \& Elitzur 1997), with the combination of
$(\tau L^{3/4})/(\psi^{1/2}\dot{M})$ being approximately constant. Indeed it
is, roughly, but with scatter ($\approx0.2$ dex; Fig.\ 7) due to slight
mis-matches between $\tau$ and $BC$ to the exact shape of the SED. The amount
of scatter is rather modest, though, and suggests that the mass-loss rates for
these stars are accurate to well within a factor two (cf.\ Srinivasan, Sargent
\& Meixner (2011); the luminosities for these stars are obtained with an
accuracy of around 10--30 per cent). The offset between the carbon stars and
M-type stars ($\approx0.6$ dex $\equiv 4\times$) is expected as this is due
primarily to the differences in opacity of the different dust species. Most --
though not all -- stars originally classified as carbon stars but fitted with
silicates end up in a location slightly off-set from that of the M-type stars.
This suggests that most of these stars are indeed different from the M-type
stars and therefore quite possibly genuine carbon stars, but that some may
have been misclassified and are in fact M type.

%
% FIGURE 8
%
\begin{figure}
\centerline{\vbox{
\psfig{figure=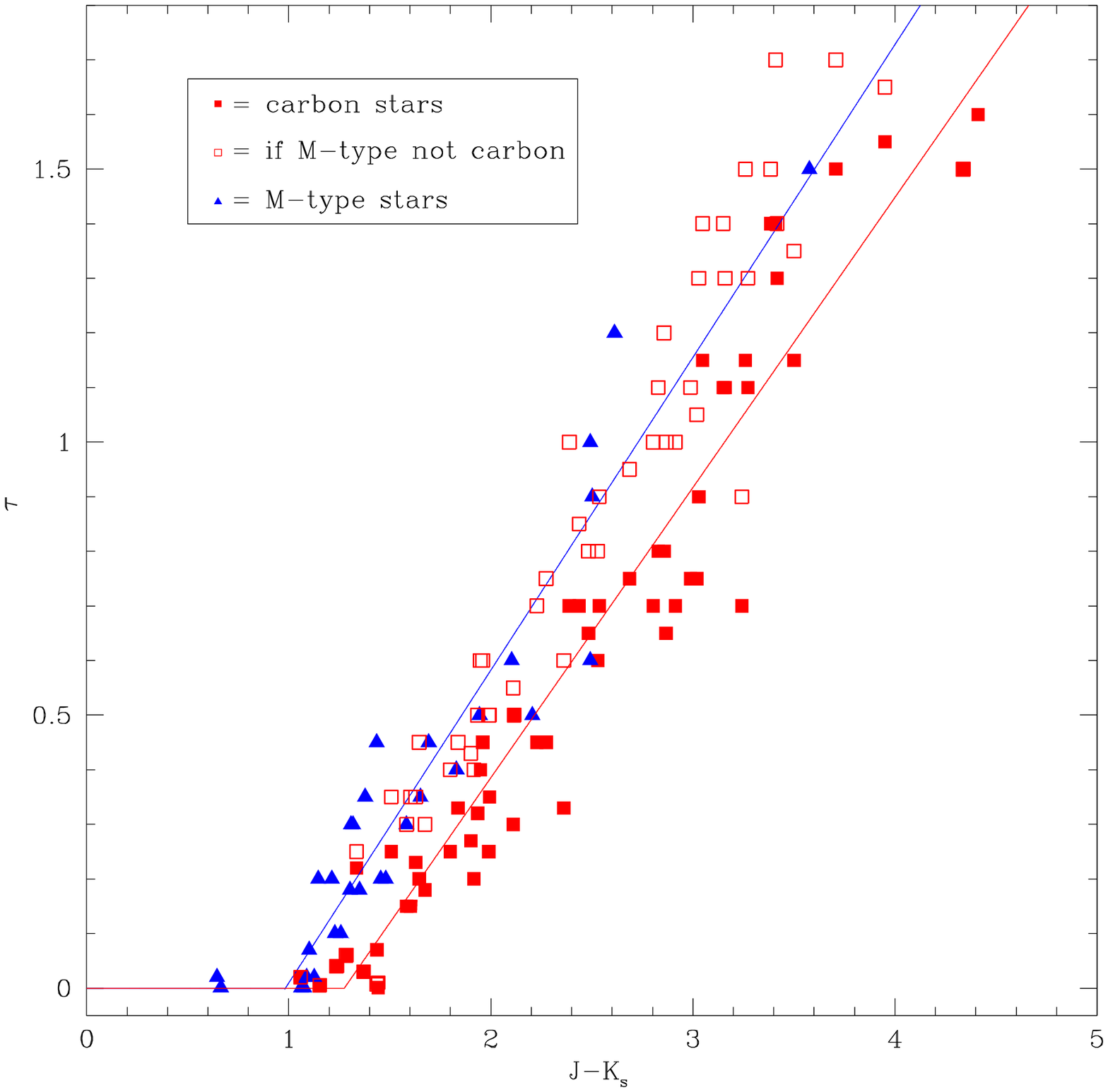,width=84mm}
\psfig{figure=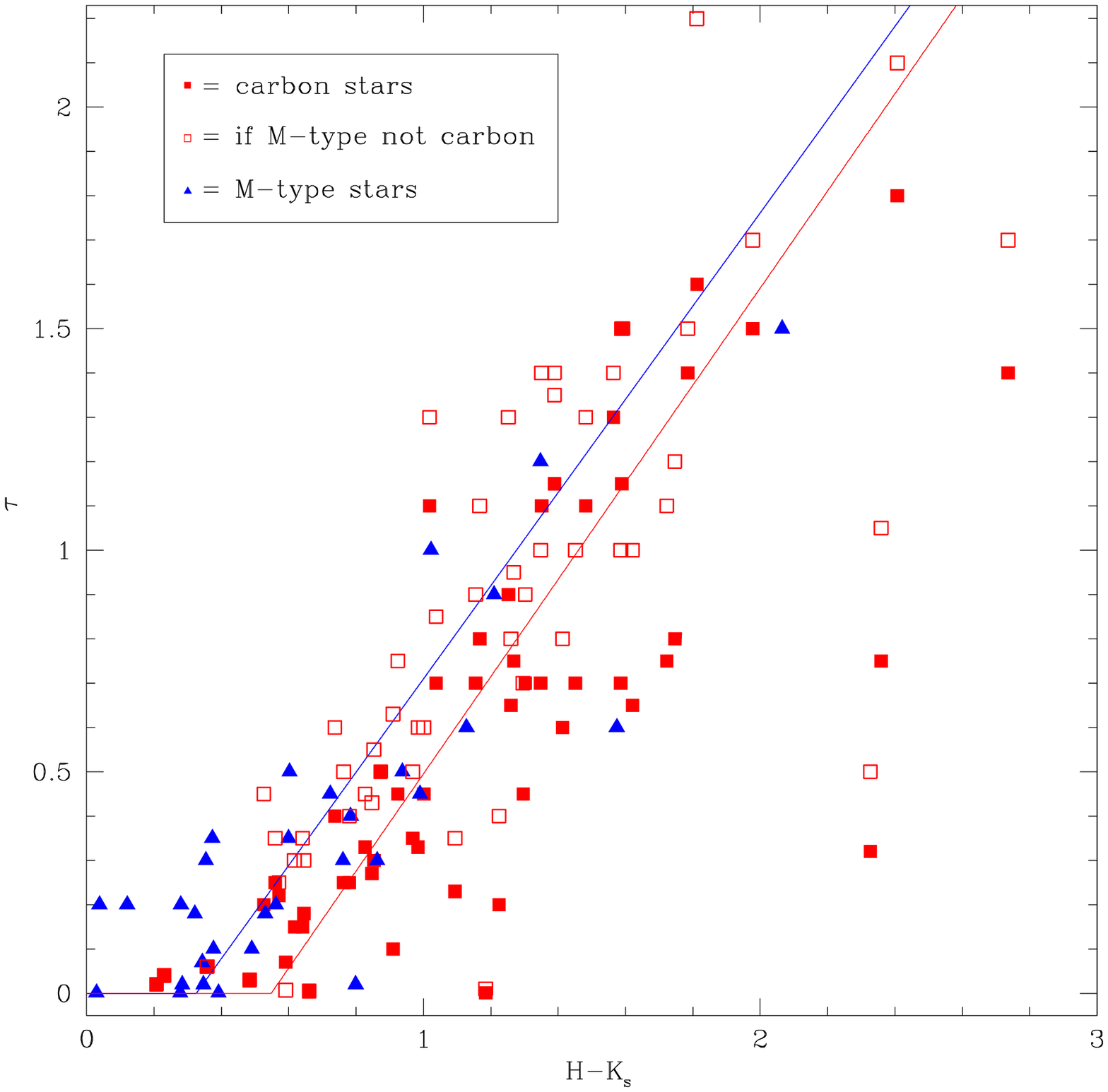,width=84mm}
}}
\caption[]{Relations between optical depth ($\tau$) and near-IR colours (top:
J--K$_{\rm s}$; bottom: H--K$_{\rm s}$) for carbon stars (red squares) and
M-type stars (blue triangles). The open red squares show the results if the
carbon stars are presumed to be oxygen-rich instead. The red and blue lines
are the adopted relations for carbon and M-type stars, respectively.}
\end{figure}

%
% TABLE 2
%
\begin{table}
\caption{Parameterisations of relations between optical depth ($\tau$) and
near-IR colour ($C$) of the form $\tau=a+b\ C$ for $C\geq c$ mag (and $\tau=0$
for $C<c$ mag).}
\begin{tabular}{llllclll}
\hline\hline
C              & a                & b     & c         &
      \mbox{ } & a                & b     & c         \\
               & \multicolumn{3}{c}{\it carbon stars} &
               & \multicolumn{3}{c}{\it M-type stars} \\
\hline
J--K$_{\rm s}$ & \llap{$-$}0.678  & 0.531 & 1.275     &
               & \llap{$-$}0.5635 & 0.573 & 0.98      \\
H--K$_{\rm s}$ & \llap{$-$}0.601  & 1.096 & 0.548     &
               & \llap{$-$}0.343  & 1.052 & 0.33      \\
\hline
\end{tabular}
\end{table}

The measured optical depth correlates with near-IR colour -- in particular the
relation with J--K$_{\rm s}$ is tight and straight (Fig.\ 8). Because the
relation between optical depth and colour directly reflects the optical
properties of the dust species, carbon stars fitted with silicates land on the
relationship for M-type stars fitted with silicates. Where J (and K$_{\rm s}$)
band photometry is available the optical depth is estimated by applying a
parameterisation of the measured relation between $\tau$ and J--K$_{\rm s}$;
if only H (and K$_{\rm s}$, but not J) band photometry is available then we
apply a parameterisation of the relation between $\tau$ and H--K$_{\rm s}$ (see
Table 2).

%
% FIGURE 9
%
\begin{figure}
\centerline{\vbox{
\psfig{figure=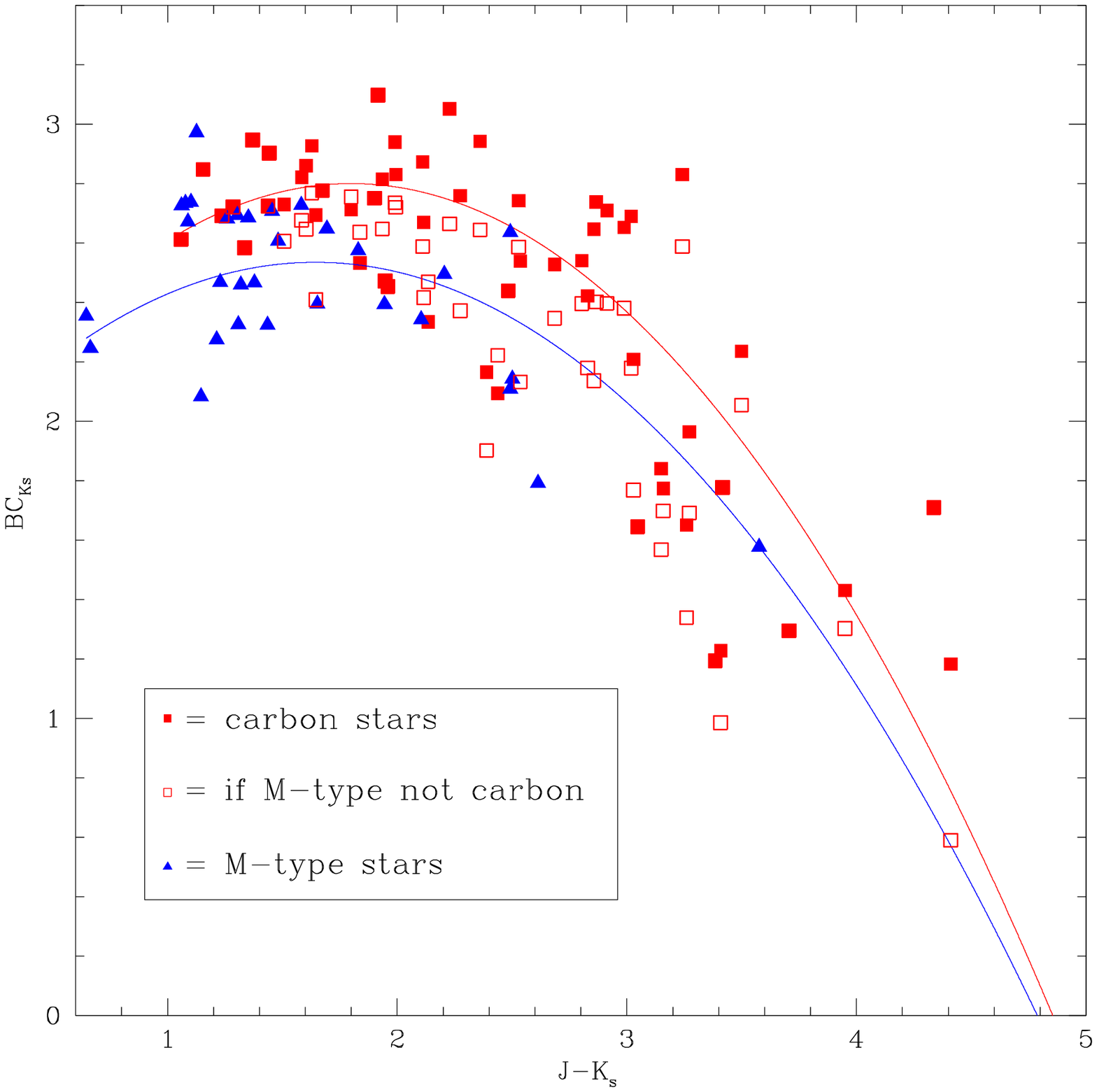,width=84mm}
\psfig{figure=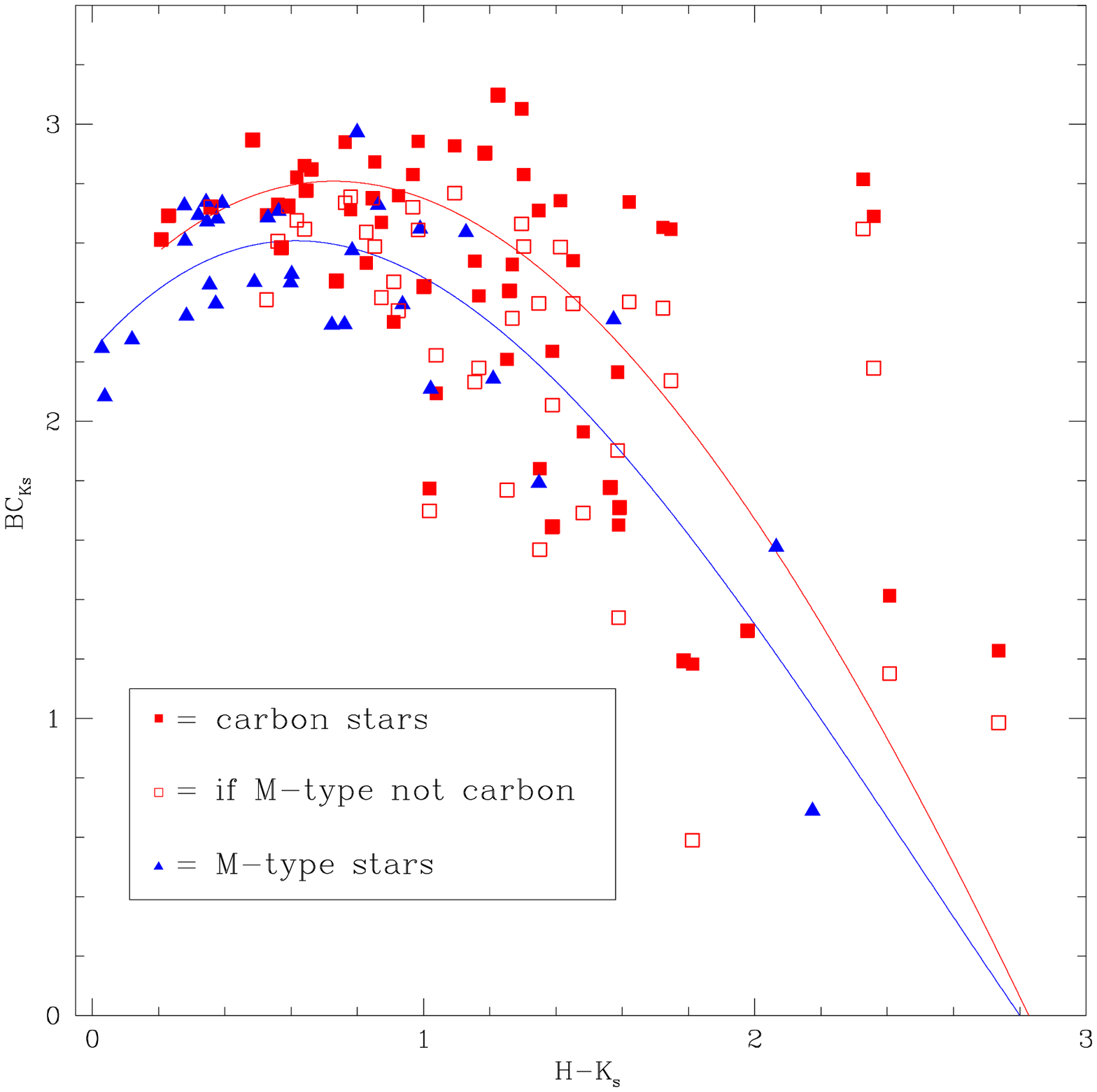,width=84mm}
}}
\caption[]{Relations between the bolometric correction to the K$_{\rm s}$ band
($BC_{\rm Ks}$) and near-IR colours (top: J--K$_{\rm s}$; bottom: H--K$_{\rm
s}$) for carbon stars (red squares) and M-type stars (blue triangles). The
open red squares show the results if the carbon stars are presumed to be
oxygen-rich instead. The red and blue lines are the adopted relations for
carbon and M-type stars, respectively.}
\end{figure}

%
% TABLE 3
%
\begin{table*}
\caption{Parameterisations of relations between the bolometric correction to
the K$_{\rm s}$ band ($BC_{\rm Ks}$) and near-IR colour ($C$) of the form
$BC_{\rm Ks}=a+b\ C+c\ C^2+d\ C^3$.}
\begin{tabular}{lclllllclllll}
\hline\hline
C       & \mbox{ } & a    & b              & c              & d    & e   &
          \mbox{ } & a    & b              & c              & d    & e   \\
                 & & \multicolumn{5}{c}{\it carbon stars}                &
                   & \multicolumn{5}{c}{\it M-type stars}                \\
\hline
J--K$_{\rm s}$   & & 1.83 & 1.08           & \llap{$-$}0.3  &      &     &
                   & 1.84 & 0.84           & \llap{$-$}0.26 &      &     \\
H--K$_{\rm s}<e$ & & 2.35 & 1.30           & \llap{$-$}0.98 & 0.08 & 5   &
                   & 2.24 & 1.27           & \llap{$-$}1.18 & 0.16 & 2.9 \\
H--K$_{\rm s}>e$ & & 6.97 & \llap{$-$}2.52 &                &      & 5   &
                   & 4.42 & \llap{$-$}1.56 &                &      & 2.9 \\
\hline
\end{tabular}
\end{table*}

Likewise, the bolometric correction to the $K_{\rm s}$ band (BC$_{\rm Ks}$)
shows a dependency on near-IR colour (Fig.\ 9) -- albeit with considerable
scatter and with occasionally large deviations for H--K$_{\rm s}$. This
relation depends on the underlying star (its temperature), and the carbon
stars fitted with silicates fall in between carbon stars fitted with
carbonaceous dust and M-type stars fitted with silicates; this also drives the
difference noted in Figure 7. Luminosities are estimated for the sample at
large by applying a parametrisation of the relation between BC$_{\rm Ks}$ and
J--K$_{\rm s}$ if $J$ is available and H--K$_{\rm s}$ if $H$ (but not $J$) is
available (see Table 3). The scatter in Fig.\ 9 suggests that an accuracy in
the luminosity is achieved of typically 30--50 per cent.

There exist pitfalls with this approach, which could lead to sources being
erroneously assigned high luminosities and mass-loss rates. One such case
comprises stars that have no J- or H-band counterpart because they are near to
the edges of the survey area, or because they are heavily extincted by
interstellar dust. Such sources would not normally have been identified as
variable in our survey. We thus decided that we would ignore stars that have
no J- or H-band magnitude and are not variable. This includes the few very
bright mid-IR sources mentioned at the end of Section 2.2, that are likely
non-stellar in origin. When they have no J- or H-band magnitude but are
variable, inspection of the brightest such sources revealed no suspicion
regarding their photometry and so for these stars we do assign a J-band
magnitude viz.\ equal to a (conservative) detection limit of $J=21$ mag. Of
just twelve such cases, the most extreme example is \#11887: a large-amplitude
variable ($A_{\rm Ks}=1.6$ mag), with luminosity $\log L=4.50$ and mass-loss
rate $\log\dot{M}=-3.9$.

%=========================================================================== 3
\section{Results}

%------------------------------------------------------------------------- 3.1
\subsection{Dust production and mass-loss rates as a function of stellar
parameters}

%
% FIGURE 10
%
\begin{figure}
\centerline{\psfig{figure=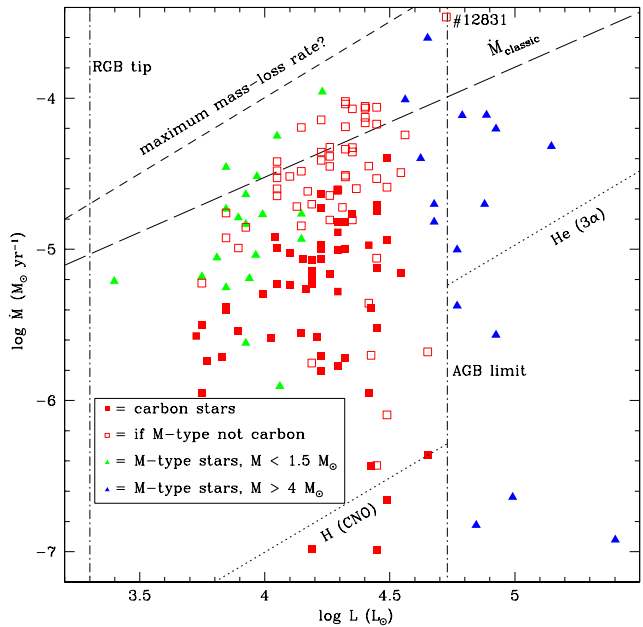,width=84mm}}
\caption[]{Mass-loss rate vs.\ luminosity, modelled with {\sc dusty}, for
low-mass AGB stars (green triangles), intermediate-mass AGB carbon stars (red
squares) and massive AGB stars and red supergiants (blue triangles). The open
red squares show the results if the carbon stars are presumed to be
oxygen-rich instead. The tip luminosity of the RGB and classical limit of the
most massive AGB stars (excluding the effects of Hot Bottom Burning) are
indicated with vertical dash--dotted lines; the mass-consumption rates by
shell hydrogen burning (CNO cycle) on the AGB and core helium burning
(triple-$\alpha$ reaction) in red supergiants are indicated with diagonal
dotted lines; and the limits to the mass-loss rate in dust-driven winds due to
single scattering (classic) and multiple scattering (maximum?) are delineated
with diagonal dashed lines (see van Loon et al.\ 1999). The highest rates are
achieved by the most massive stars, but there is significant spread at all
luminosities suggestive of evolutionary effects.}
\end{figure}

The mass-loss rate shows some dependence on luminosity (Fig.\ 10). Firstly,
the highest rates are generally achieved by the most luminous, most massive
stars, in agreement with earlier findings in the Magellanic Clouds (van Loon
et al.\ 1999; Srinivasan et al.\ 2009). Except for a few more extreme examples
that would probably have been missed in our survey, carbon stars in the Milky
Way reach mass-loss rates of a few $\times10^{-5}$ M$_\odot$ yr$^{-1}$
(Whitelock et al.\ 2006), whilst those in the Large Magellanic Cloud (LMC)
reach $\dot{M}\sim10^{-5}$ M$_\odot$ yr$^{-1}$ (Gullieuszik et al.\ 2012), i.e.\
comparable to the most extreme carbon stars in our M\,33 sample. In the Solar
Neighbourhood, M-type AGB stars reach similar mass-loss rates, a few
$\times10^{-5}$ M$_\odot$ yr$^{-1}$ (Jura \& Kleinmann 1989), whilst M-type
supergiants display rates between $10^{-7}<\dot{M}<10^{-4}$ M$_\odot$ yr$^{-1}$
with the majority in excess of a few $\times10^{-6}$ M$_\odot$ yr$^{-1}$ (Jura
\& Kleinmann 1990), again both very similar to what we find in M\,33.

Much of the spread in mass-loss rate at a given luminosity is likely to
reflect stellar evolution (cf.\ van Loon et al.\ 1999, 2005a). The least
luminous carbon stars ($\log L<4$) are probably not yet at the tip of their
AGB because M-type AGB stars are found at the same luminosities but higher
mass-loss rates -- the latter are likely more evolved than the carbon stars;
but these carbon stars are not expected to become M-type again (Marigo et al.\
2008) so they must still evolve to higher luminosities before they end their
AGB evolution. With $\dot{M}$ of a few $\times10^{-6}$ M$_\odot$ yr$^{-1}$ they
exhibit several times lower mass-loss rates than those near the tip of their
AGB evolution (around $\log L\sim4.3\equiv 20,000$ L$_\odot$) where they reach
$\dot{M}$ in excess of $10^{-5}$ M$_\odot$ yr$^{-1}$. This suggests that the
mass-loss rate increases as a star climbs the AGB.

Alternatively, the less luminous carbon stars with lower mass-loss rates may
be in their inter-thermal pulse luminosity dip and/or the more luminous carbon
stars with high mass-loss rates may just be experiencing the aftermath of a
thermal pulse (Olofsson et al.\ 1990; Vassiliadis \& Wood 1993; Mattsson,
H\"ofner \& Herwig 2007).

Fitting the stars that we originally classified as carbon stars with silicates
generally yields higher mass-loss rates because of the lower specific opacity
of silicates compared to amorphous carbon grains. One star, \#12831
($J-K_{\rm s}=4.4$ mag, $A_{\rm Ks}=0.63$ mag) has in this way become one of the
more extremely mass-losing objects, with $\log\dot{M}=-3.5$ nine times higher
than if fitted with silicates. The inferred birth mass would now be 6.7
M$_\odot$.

Evolution of red supergiants is mostly in terms of effective temperature
($T_{\rm eff}$) rather than luminosity. The dusty wind mass-loss rate depends
sensitively on $T_{\rm eff}$ (van Loon et al.\ 2005a; cf.\ Bonanos et al.\
2010); because red supergiants do not become as cool as massive AGB stars the
highest rates achieved by red supergiants fall a little behind those achieved
by the most extreme AGB stars considering that the red supergiants are more
luminous (Fig.\ 10). Yet there is no luminosity gap in the stars with the
highest mass-loss rates, confirming our earlier suggestion (Paper II) that
super-AGB stars become dust-enshrouded. Super-AGB stars are strictly speaking
red supergiants. However, the most massive AGB stars -- i.e.\ those that will
{\it not} ignite core carbon burning -- experience Hot Bottom Burning (HBB;
Iben \& Renzini 1983); this not only prevents them from turning into carbon
stars but also enhances their luminosity up and above the classical
core-mass--luminosity relation (Boothroyd \& Sackmann 1992). These stars may
thus reach luminosities that exceed the classical AGB limit (indicated in
Fig.\ 10), making it difficult to distinguish between an AGB star experiencing
HBB and a super-AGB star or red supergiant. They may separate in
$T_{\rm eff}$--$L$ diagrams (van Loon et al.\ 2005a) and thus also in $P$--$L$
diagrams (cf.\ Wood et al.\ 1992; Whitelock et al.\ 2003) and perhaps in
$\dot{M}$--$L$ diagrams too. It is not very well established what is the
maximum luminosity than can be reached under the influence of HBB, but it
seems unlikely that it comprises much more than 50 per cent (0.2 dex).
Depending on the timing of the ensuing supernova explosion, red supergiants
might be found in the dust-enshrouded phase, or a preceding or following phase
characterised by a thinner dust envelope. But if super-AGB stars explode they
are likely to be dust-enshrouded, explaining explosions such as SN\,2008S
(Botticella et al.\ 2009).

%
% FIGURE 11
%
\begin{figure}
\centerline{\vbox{
\psfig{figure=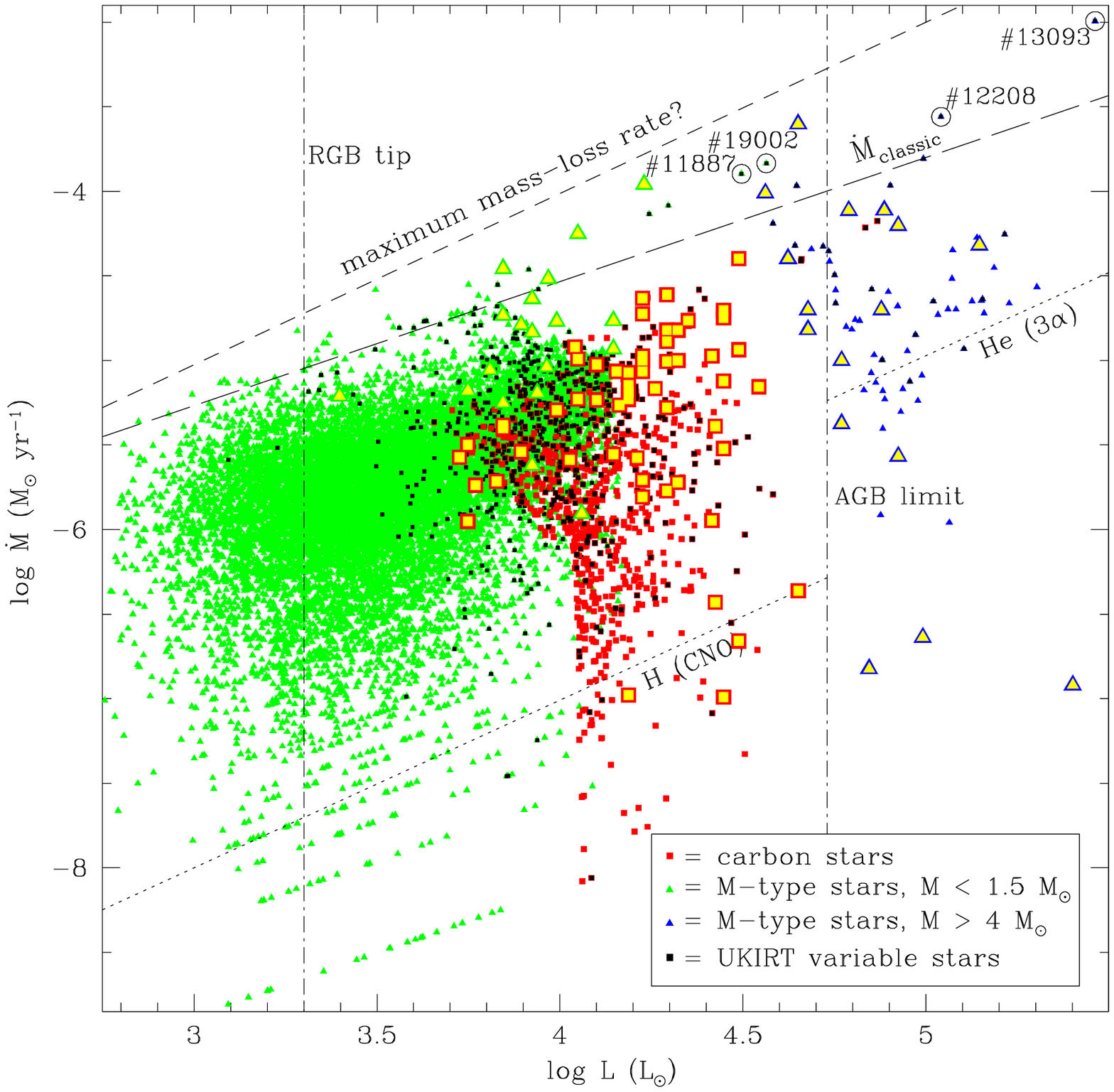,width=84mm}
\psfig{figure=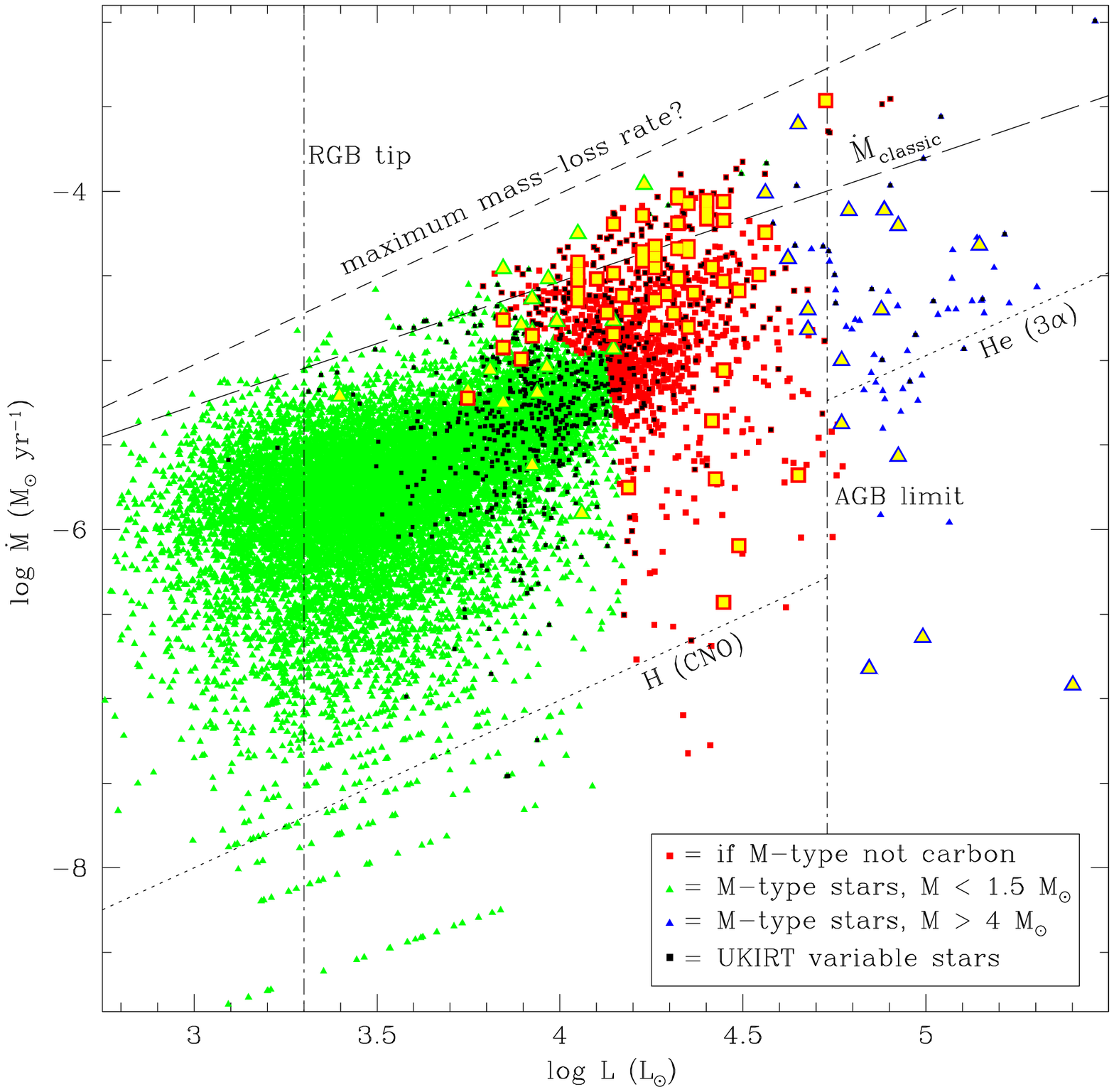,width=84mm}
}}
\caption[]{Mass-loss rate vs.\ luminosity. Massive, luminous M-type stars are
depicted by blue triangles; AGB carbon stars by red squares; and low-mass
M-type stars (at lower luminosities) by green triangles. Large yellow symbols
identify the stars modelled with {\sc dusty}; other UKIRT variable stars are
identified by black squares. The most extreme mass-losing stars are labelled.
The bottom panel shows the results if the carbon stars are presumed to be
oxygen-rich instead.}
\end{figure}

The sample of stars that were modelled with {\sc dusty} is not complete, and
we therefore include in Figure 11 also the UKIRT variable stars for which no
{\it Spitzer} photometry is available as well as non-variable stars. We note
the most extreme datum in this graph, \#13093, at $\log L=5.46$ and
$\log\dot{M}=-3.0$. With $K_{\rm s}=16.96$ mag and $J-K_{\rm s}=5.35$ mag it
is consistent with being a heavily-reddened $\simeq14$ M$_\odot$ supergiant;
we also identified it as a large-amplitude variable ($A_{\rm Ks}=1.05$ mag).
Our K$_{\rm s}$-band image shows a point source unaffected by crowding. No
{\it Spitzer} photometry is available, neither from McQuinn et al.\ (2007) nor
Thompson et al.\ (2009). Inspection of the IRAC images (3.6, 4.5, 5.8 and 8
$\mu$m) from the {\it Spitzer} archive, however, clearly reveals the source,
adding credence to it being an extremely dusty evolved star. It sits next to
an extended area of mid-IR emission, though, which may explain why it was not
measured. It is almost twice as bright at 8 $\mu$m as the highest mass-losing
UKIRT variable that we modelled with {\sc dusty} (\#15552), broadly in line
with the factor 2.5 difference in their mass-loss rates.

On the other hand, the star with the next-highest mass-loss rate, \#12208
($\log L=5.04$, $\log\dot{M}=-3.6$) is more suspicious. With $K_{\rm s}=15.3$
mag it is bright, and with $J-K_{\rm s}=3.4$ mag it is red, but its estimated
K$_{\rm s}$-band amplitude is moderate ($A_{\rm Ks}=0.35$ mag). It is the
brightest UKIRT variable in this central region of M\,33 at 8 $\mu$m, by far
(see Fig.\ 4). While it appears unresolved on the 8-$\mu$m image it sits on
top of more complex emission, within $6^{\prime\prime}$ of a catalogued
molecular cloud (Bolatto et al.\ 2008) and a supernova remnant (Gordon et al.\
1999); this suggests it might be a young stellar object or even an
ultra-compact H\,{\sc ii} region.

A third object worth noting is the candidate $\eta$\,Carin{\ae} analogue
M\,33-8 (Khan, Stanek \& Kochanek 2012). It is the only such object in their
list of nine in M\,33 that (just) falls within our UIST/UFTI survey. They
combined optical photometry with 2MASS near-IR and {\it Spitzer} mid-IR data,
to derive the spectral energy distribution and luminosity. They show a {\it
Hubble} Space Telescope image in the visual dominated by a single source
surrounded by a clustering of fainter stars. Our UFTI image reveals the
dominant source is in fact a duplet in the K$_{\rm s}$ band, which would appear
unresolved in the 2MASS image. The UIST catalogue lists them as \#18469, with
$K_{\rm s}=17.75$ mag, and \#18441, with $K_{\rm s}=17.89$ mag. This compares
well with the I-band magnitude (0.8 $\mu$m) listed in Khan et al.\ (2012) of
$I=17.68$ mag. We estimate luminosities of $\log L=3.88$ and 3.68,
respectively; i.e.\ together they fall well short of accounting for the
integrated IR luminosity of $\log L_{\rm IR}=5.56$ derived by Khan et al. The
latter must include significant contributions from other sources, especially
as it includes the 24-$\mu$m {\it Spitzer} emission which has an angular
resolution of only $6^{\prime\prime}$. We derive mass-loss rates of
$\log\dot{M}=-5.2$ and $-5.1$ for \#18469 and \#18441, respectively;
substantial, but not extreme.

Close pairs of infrared-bright stars are not unique. For instance, \#19002 is
an unresolved blend of two roughly equally-bright stars. We assigned a
luminosity of $\log L=4.56$ and $\log\dot{M}=-3.8$ to the blended object;
certainly at least one of the two stars is very dusty.

%
% FIGURE 12
%
\begin{figure}
\centerline{\vbox{
\psfig{figure=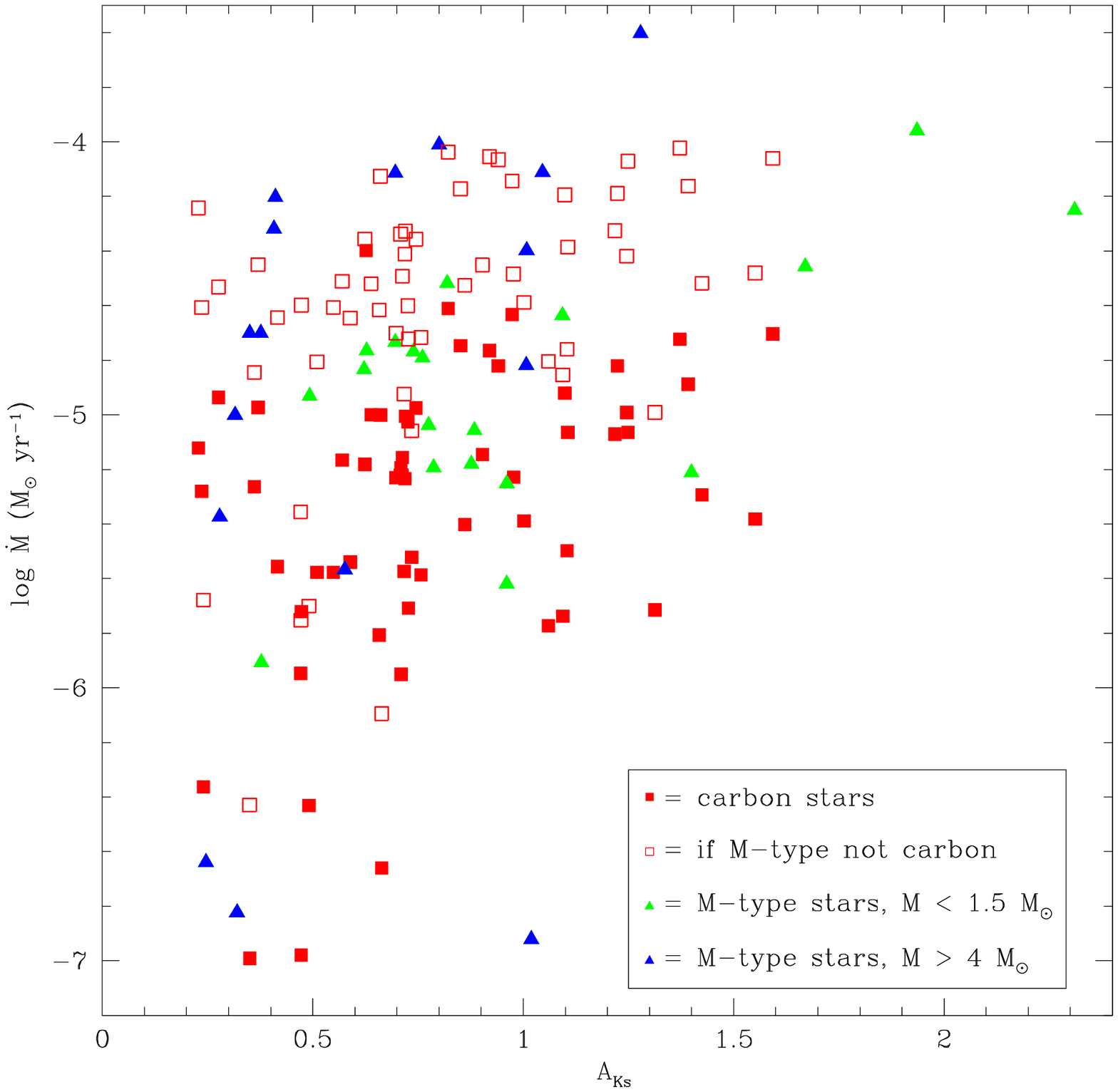,width=84mm}
\psfig{figure=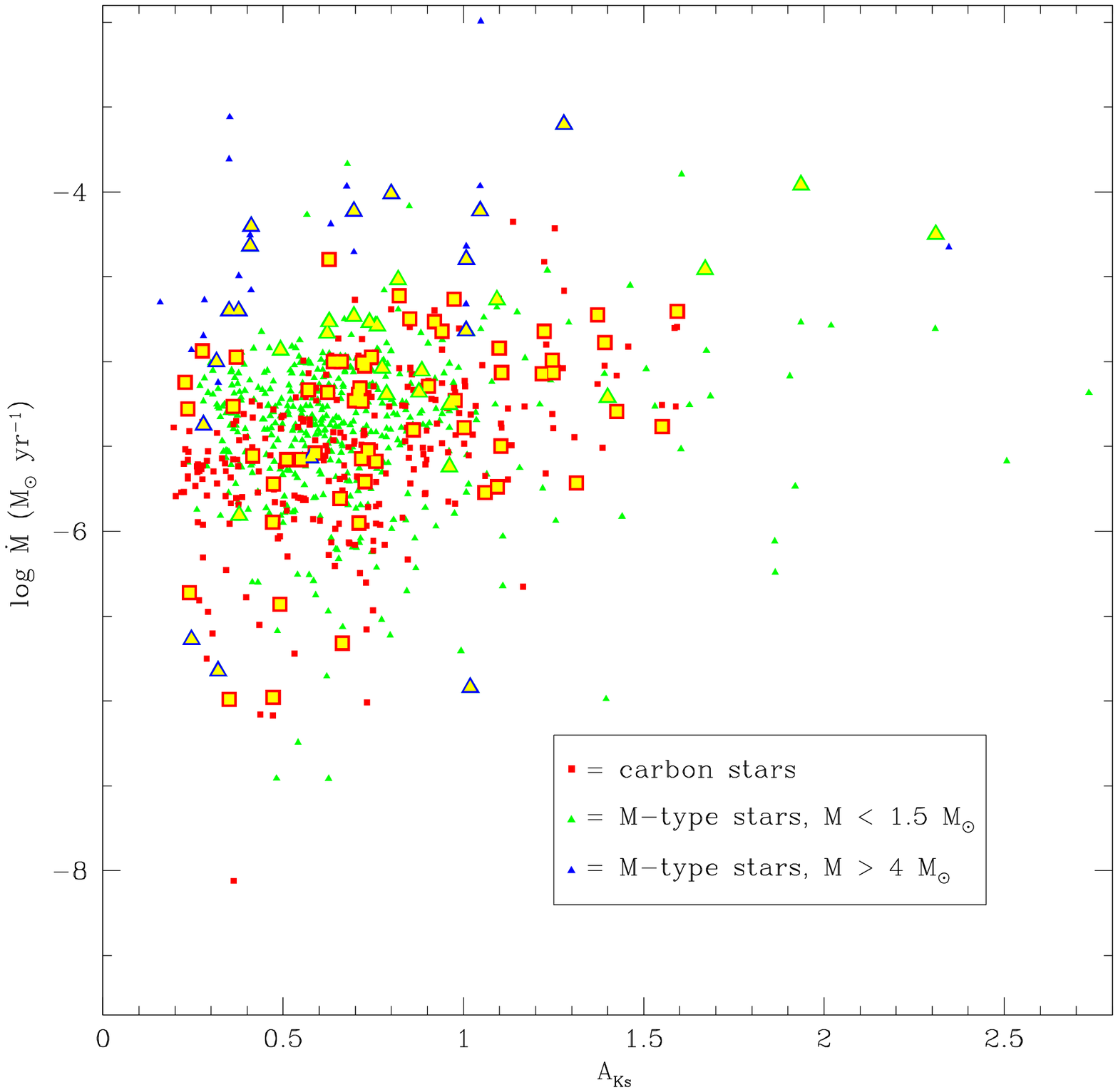,width=84mm}
}}
\caption[]{Mass-loss rate vs.\ K$_{\rm s}$-band amplitude. Top: stars modelled
with {\sc dusty} -- the open red squares show the results if the carbon stars
are presumed to be oxygen-rich instead; bottom: all UKIRT variable stars,
where large yellow symbols identify the stars modelled with {\sc dusty}.}
\end{figure}

The mass-loss rate increases with increasing K$_{\rm s}$-band amplitude (Fig.\
12). The correlation is quite clear in spite of the relative inaccuracy of the
amplitudes (determined from sparsely sampled lightcurves; see Paper I).
 The correlation is consistent with a r\^ole for stellar pulsation in
driving the winds from cool stars;
it also confirms that, whilst the {\it relative} amplitudes (expressed in
magnitudes) are smaller for more luminous stars (the M-type stars with $M>4$
M$_\odot$ are the most luminous, those with $M<1.5$ M$_\odot$ are generally the
least luminous), for the same amplitude the mass-loss rates are higher for
more luminous stars: mass loss is more directly related to the {\it absolute}
amplitude (expressed in luminosity units) (van Loon et al.\ 2008). The
K$_{\rm s}$ band is relatively insensitive to variations in stellar
temperature, circumstellar extinction and dust emission as it is near the peak
of the stellar SED and in between the strong attenuation by dust at shorter
wavelengths and its emission at longer wavelengths -- hence also the
preference for expressing bolometric corrections in relation to the K or
K$_{\rm s}$ band.
 However, it is important, but difficult in practice, to separate the
effects of pulsation and temperature on mass loss, as both the pulsation
strengthens and the temperature decreases as the star ascends the RGB, AGB or
red supergiant branch (see Whitelock, Feast \& Pottasch 1987).
The carbon stars confirm this picture in the sense that their amplitudes are
intermediate between the more and less luminous M-type stars, but their
mass-loss rates seem a little lower. As expected, the mass-loss rates of
presumed carbon stars when fitted with silicates fall in between those of
low-mass and higher-mass M-type stars. This fact neither confirms nor refutes
their carbon star nature.

%
% FIGURE 13
%
\begin{figure}
\centerline{\vbox{
\psfig{figure=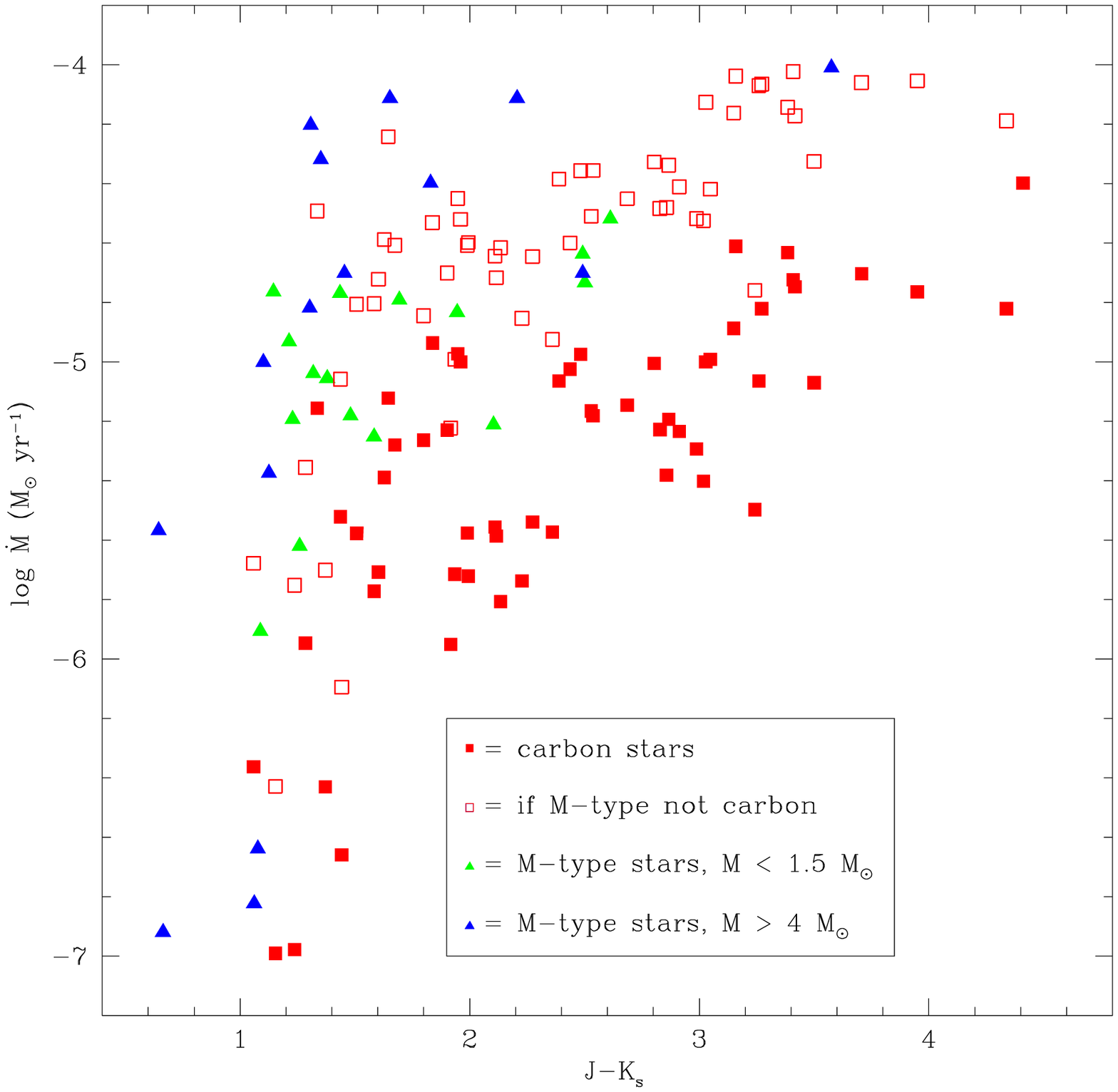,width=84mm}
\psfig{figure=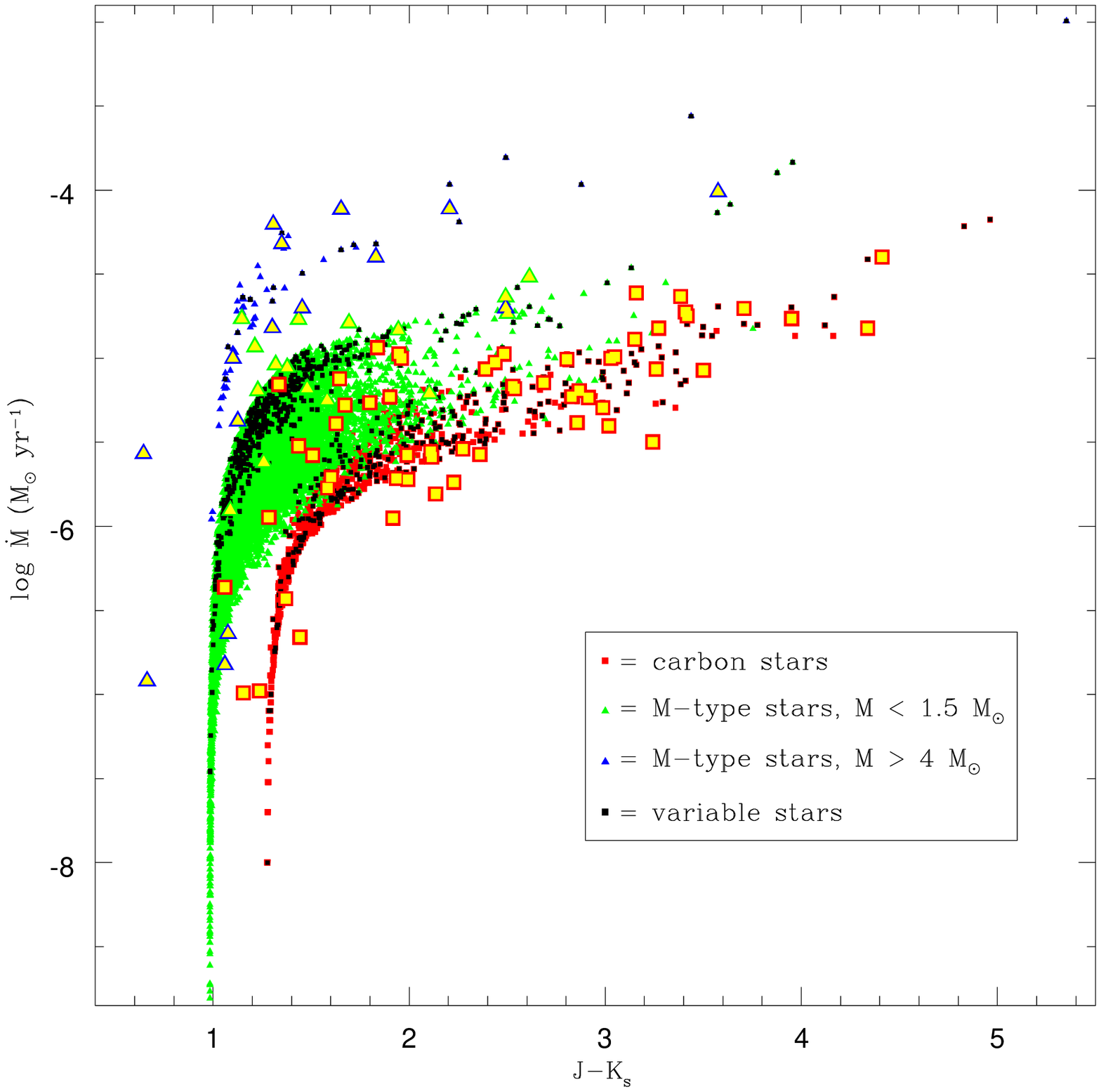,width=84mm}
}}
\caption[]{Mass-loss rate vs.\ J--K$_{\rm s}$ colour. Top: stars modelled with
{\sc dusty} -- the open red squares show the results if the carbon stars are
presumed to be oxygen-rich instead; bottom: all stars, where large yellow
symbols identify the stars modelled with {\sc dusty} and little black symbols
identify the remaining UKIRT variable stars.}
\end{figure}

Mass-loss rates have often been correlated with infrared colour, to provide a
simple recipe for converting infrared colours to mass-loss rates; however,
these methods depend on the star's luminosity as well as the dust properties
(see van Loon 2007 for a review). Figure 13 exemplifies this once more: the
carbon stars become easily reddened by their dust but that does not
necessarily imply as high a mass-loss rate as a less reddened M-type star.
Likewise, among stars with similar dust the less luminous stars are more
readily reddened than more luminous ones (van Loon et al.\ 1999). Thus, care
must be taken when converting an infrared colour into a mass-loss rate. The
carbon stars' luminosities are intermediate between those of the low-mass and
higher-mass M-type stars, and indeed their mass-loss rates fall in between
those groups' mass-loss rates if also the presumed carbon stars are fitted
with silicates.

%------------------------------------------------------------------------- 3.2
\subsection{Feedback into the interstellar medium}

Assessments of the integrated mass loss (or dust production) from a stellar
population are fraught with uncertainties. In small populations, stochastic
effects are caused by very small numbers of stars which may dominate the
budget (van Loon et al.\ 2005b; McDonald et al.\ 2009), but it is also
difficult to accurately account for the many stars that each contribute
little. Most red giants undergo mild mass loss and their circumstellar
envelopes are not very dusty; photometric tracers of this mass loss and dust
will be affected by photometric uncertainties and scatter in interstellar
reddening, as well as uncertainties in the underlying photospheric spectral
energy distribution (cf.\ McDonald, Zijlstra \& Boyer 2012). Interstellar
reddening towards the central region of M\,33 is modest (see Paper I); whilst
it is negligible for the dustiest stars it may lead to an over-estimate of the
amount of dust produced by stars with little mass loss. Photometric scatter
will also cause some stars to appear to have more dust than they really do,
but equally cause the opposite effect in others. As this is of a purely
statistical nature we can try to compensate for it by considering the
distribution of stars over negative values for the optical depth as one would
have derived them from the near-IR colour relation (Fig.\ 14, top). A similar
distribution of erroneous optical depth values would extent towards positive
values, and this distribution can be subtracted to yield the true distribution
over optical depth (Fig.\ 14, bottom).

%
% FIGURE 14
%
\begin{figure}
\centerline{\psfig{figure=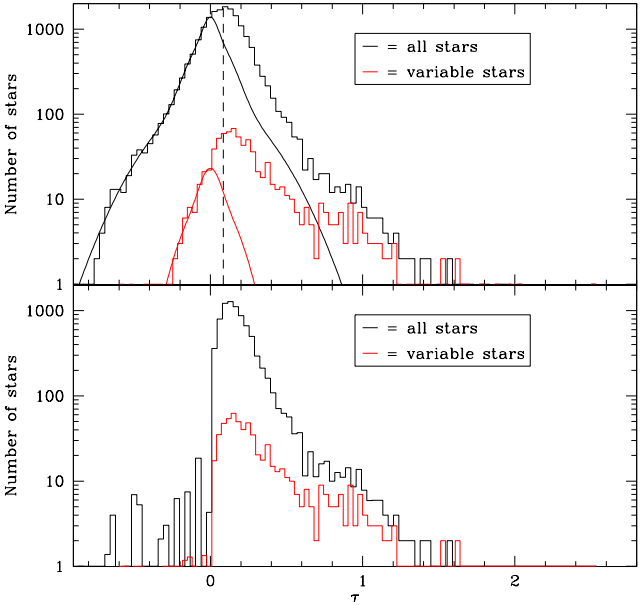,width=84mm}}
\caption[]{Top: Distribution of stars over optical depth, $\tau$, and bottom:
after subtracting a multi-Gaussian fit to the negative values of $\tau$ which
was centered at $\tau=0$. The UKIRT variables are plotted also separately, in
red. The vertical dashed line marks the peak of the distribution of all
stars.}
\end{figure}

The fraction of stars that were identified to be variable increases with
increasing optical depth, to about unity for the most obscured specimens
(Fig.\ 14). This is encouraging, suggesting that the dustiest stars produced
most of that dust themselves. We cannot be certain, however, that the dust
inferred for the non-variables and/or less obscured stars was all produced by
themselves.

Because the mass-loss and dust-loss rate is derived from the optical depth via
scaling with the luminosity, correcting the mass return is not
straightforward. This correction can be applied statistically in two ways: (1)
by reducing the contribution from each star by a factor $f/f_0$ where $f_0$
and $f$ are the distributions before and after correcting for the symmetric
distribution around $\tau=0$, where $f/f_0\equiv 0$ for $\tau<0$; or (2) by
including all stars including those with $\tau<0$, and assigning negative
mass-loss rates to the latter. The problem with the second method is that some
stars with very negative values and/or very high luminosities could make a
large difference to the end result, whereas in the first method the correction
is ``smoother''. On the other hand, the second method accounts, to some
extent, for variations of the correction factor with luminosity should such
dependence exist. In what follows we have adopted the second method.

%
% FIGURE 15
%
\begin{figure*}
\centerline{\vbox{
\hbox{\psfig{figure=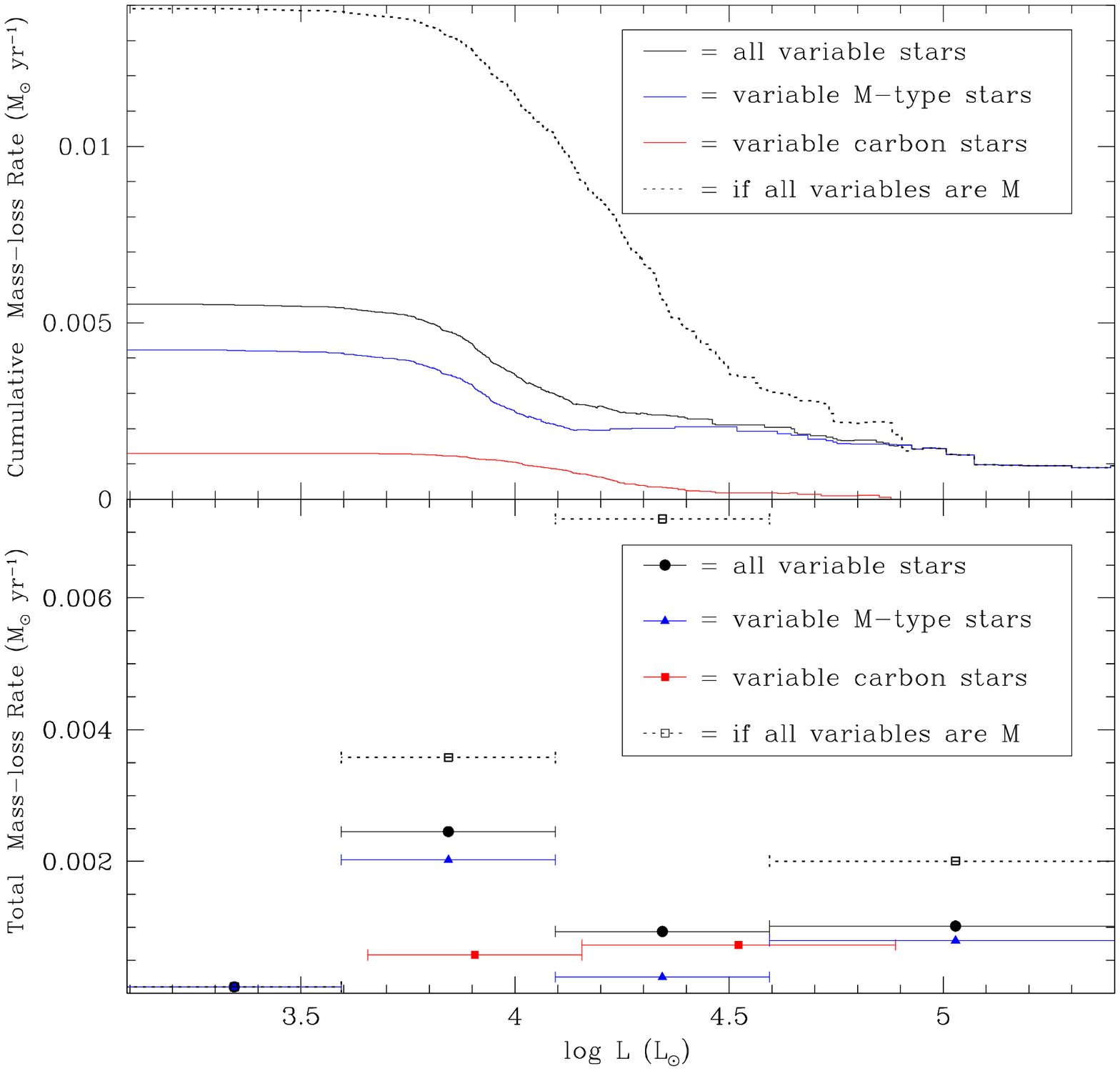,width=88mm}
\psfig{figure=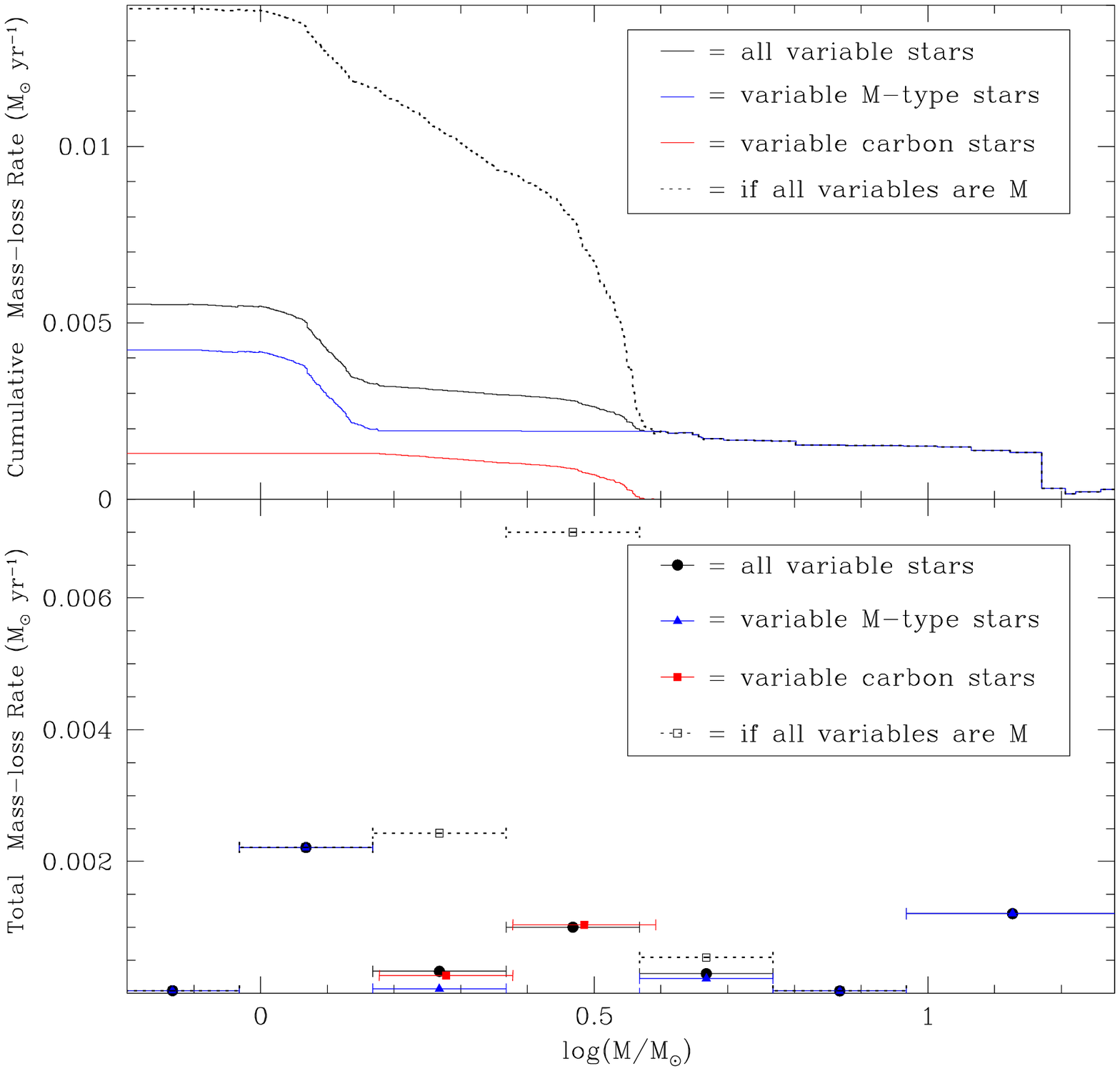,width=88mm}}
\hbox{\psfig{figure=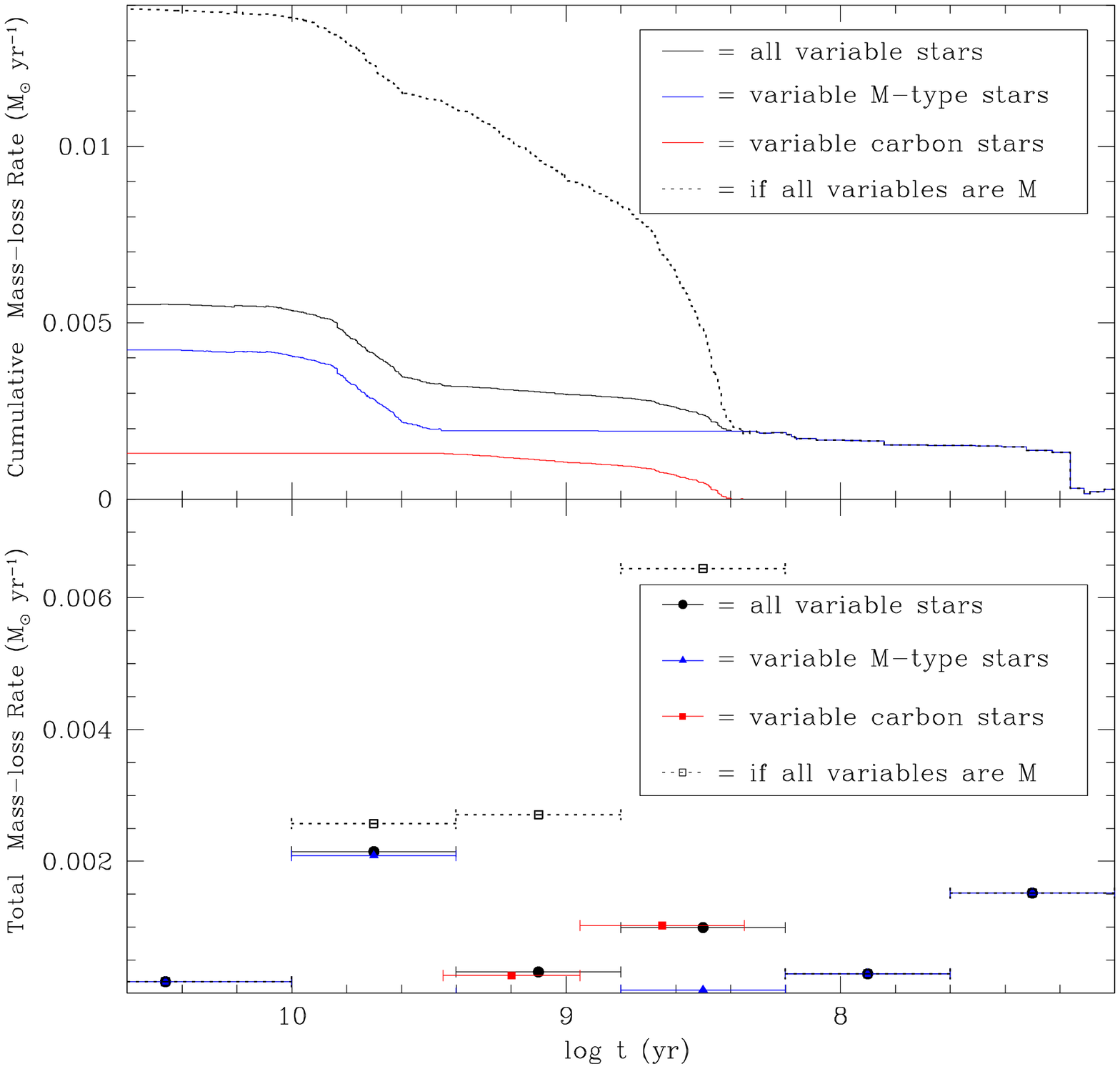,width=88mm}
\psfig{figure=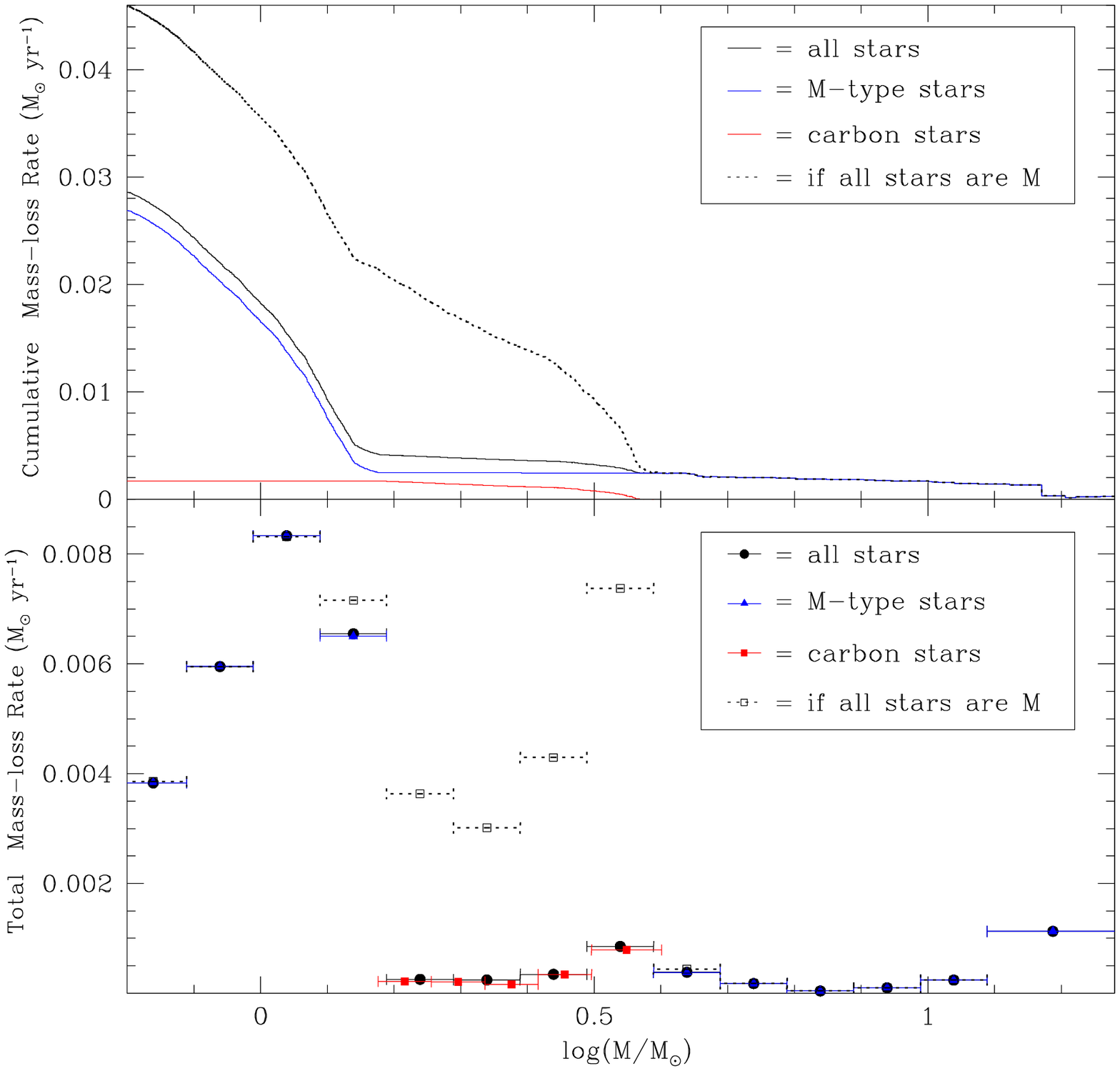,width=88mm}}
}}
\caption[]{Mass-loss rate vs.\ luminosity (top left), age (bottom left) and
birth mass (top right) for the UKIRT variable stars, and vs.\ birth mass for
all stars including non-variable stars (bottom right). The top panels show the
cumulative distribution (integrated from the right); the bottom panels show
binned contributions. Distinction is made between likely carbon stars (red)
and M-type stars (blue).}
\end{figure*}

Hence we obtain the binned and cumulative mass-loss rate distributions
depicted in figure 15. We concentrate on the mass-loss rates of the UKIRT
variable stars, as these are believed to contribute the majority of the mass
loss, but we also show the result including the non-variable stars -- which
are dominated by the low-mass red giants. The latter are numerous and
therefore appear to contribute several times as much as the intermediate-mass
AGB stars and higher-mass red supergiants, but it must be remembered that
interstellar dust will be more often the cause of their generally modest
reddening and so that result must be regarded as a firm upper limit to the
mass return, of $<0.028$ M$_\odot$ yr$^{-1}$ (or $<0.046$ M$_\odot$ yr$^{-1}$ if
all carbon stars are actually M-type stars). The true mass-return rate must be
closer to that derived from the UKIRT variable stars, $\approx0.0055$
M$_\odot$ yr$^{-1}$ (or $0.0139$ M$_\odot$ yr$^{-1}$ if all carbon stars are
actually M-type stars). Accepting that some of the mass loss might have been
missed -- as not all variable stars may have been recognised in our survey --
we could conceive a total mass-return rate of $\sim0.006$--0.01 M$_\odot$
yr$^{-1}$ (up to $\sim0.02$ M$_\odot$ yr$^{-1}$ if a large fraction of the
carbon stars are actually M-type stars). Given the assumptions and
uncertainties inherent to the mass-loss determinations we could hardly claim a
higher degree of accuracy.

% From 18398 stars the mass loss rate is just derived for 17651 stars,  which
% 16574 stars are M-type stars and 1077 stars are carbon stars (The
% classification is done based on mass of stars, supposed those with $M
% \approx 1.5-4$ are carbon stars and the others are M-type stars.). We
% excluded 747 stars from further analysis because these are non-variables
% stars with detection in just K-band which observed less than three times, so
% we believe their photometry should be poor.

That said, low-mass M-type AGB stars appear to make a similar contribution to
the mass return in the central regions of M\,33 as all more massive stars
combined. Among the latter, the massive carbon stars and red supergiants make
similar contributions, while the low-mass carbon stars and massive AGB stars
contribute less. As a result, the present-day mass-return arises mainly from
stars formed in one of three major episodes: $t<40$ Myr ago ($\log t<7.6$),
$t\sim0.2$--1 Gyr ago ($\log t\sim8.6$) and $t\sim3$--10 Gyr ago ($\log
t\sim9.7$). This mainly reflects the star formation history (Paper II). The
contribution from the intermediate-age population, formed $t\sim0.2$--1 Gyr
ago, is larger if the presumed carbon stars are actually mostly M-type stars,
in which case they account for about half of the total mass return.

It is worthy of note that carbon grains make up $<23$\% of the present-day
dust-mass return, i.e.\ the interstellar dust is predominantly oxygen-rich.
This fraction becomes even smaller if the population of carbon stars was
overestimated.

The radial profile of the mass return rate (Fig.\ 16, in terms of surface
density deprojected onto the galaxy plane -- see Paper II) is flat within the
inner $r<0.5$ kpc for the carbon stars. There appears to be a clear radial
gradient for the M-type stars, with the mass return greatest within the
central few hundred pc, but large deviations occur around $r\sim0.3$--0.4 kpc.
A more complete picture is obtained from the full 2-D map of mass return
(Fig.\ 17, {\it not} deprojected), where a smooth level of gas and dust
injection underlies dramatic local enhancements due to the concentration of
red supergiants (and super-AGB stars) near to their sites of formation in
discrete molecular cloud complexes. One of these includes \#13093 and \#12208,
the supergiants with the highest mass-loss rates (see Section 3.1). For
comparison we show a {\it Spitzer} composite in the left panel of figure 17 --
note the bright source at ($1^{\rm h}34^{\rm m}00^{\rm s}$,
$+30^\circ40^\prime47^{\prime\prime}$): this is M\,33-8 (Kahn et al.\ 2012; see
Section 3.1). The spatial distribution of mass return does not change markedly
if the carbon stars turn out to be mostly M-type stars.

%
% FIGURE 16
%
\begin{figure}
\centerline{\psfig{figure=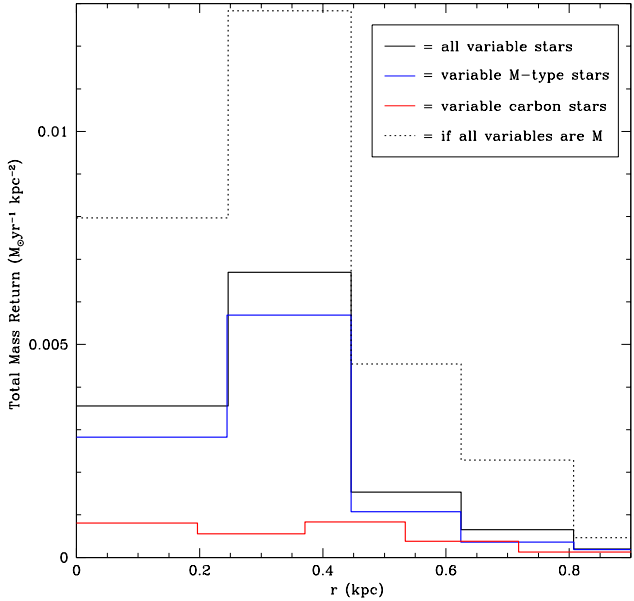,width=83mm}}
\caption[]{Radial distribution of mass-return-rate surface density over the
central region of M\,33.}
\end{figure}

%=========================================================================== 4
\section{Discussion and conclusions}

If we believe the model predictions for the duration of the pulsation phase
($\delta t$, see table 2 in Paper II and Marigo et al.\ 2008) then we can
calculate the mass lost during the pulsation phase as a fraction of the birth
mass if we sample this phase both randomly and sufficiently:
\begin{equation}
\eta = \frac{\sum_{i=1}^{N}{\left(\dot{M}_i\times(\delta t)_i\right)}}
            {\sum_{i=1}^{N}{M_i}}.
\end{equation}
Obviously, this ratio should be less than unity. It is therefore somewhat
disconcerting to find that this is not the case (Fig.\ 18), though it does
lend support to our belief that the non-variable stars cannot contribute much
to the total mass-loss budget, or it would seriously aggravate the situation.
The integrated mass-loss along the evolution of low- and intermediate-mass
stars is strongly constrained by the initial--final mass relation determined
from white dwarfs in stellar clusters. The ratio of our values to those
derived from the initial--final mass relation from Williams, Bolte \& Koester
(2009) varies between $<2$ and $\sim20$. For massive stars the situation is
more uncertain; while most of them (at the lower mass end) will leave behind
a neutron star (of order a tenth of their birth mass) it is unclear how much
mass is lost as a red supergiant, as a blue supergiant, and during the
supernova explosion (see, e.g., Ekstr\"om et al.\ 2012). But what causes the
discrepancy? Have the mass-loss rates been over-estimated? Or is the pulsation
duration in error?

If the pulsation duration was over-estimated, then the star formation rate
would have been under-estimated. At the time, we compared our estimate of the
recent star formation rate, $\psi\approx0.004$--0.007 M$_\odot$ yr$^{-1}$
kpc$^{-2}$, with work by Wilson, Scoville \& Rice (1991) who had derived
$\psi\sim0.01$ M$_\odot$ yr$^{-1}$ kpc$^{-2}$. We were satisfied that, within
the inherent uncertainties, these two estimates were consistent. In light of
the above discrepancy between birth mass and lost mass, though, we now revisit
these estimates:
\begin{itemize}
\item[1.]{It is reasonable to assume that star formation occurs within the
gas-rich disc, and so the recent star formation rate ought to be deprojected
onto the galaxy's plane. This depresses the rate by a factor $\cos i$ where
$i$ is the angle of inclination of the normal to the M\,33 disc with respect
to the line-of-sight, for which we take $i=56^\circ$ (Zaritsky et al.\ 1989).
The fact that the survey area is square but the position angle is $PA=23^\circ$
matters, but not a lot. So our estimate for the recent star formation rate
becomes $\psi\approx0.003$ M$_\odot$ yr$^{-1}$ kpc$^{-2}$.}
\item[2.]{Several other groups have estimated star formation rates for the
central region of M\,33 (as part of larger surveys), from H$\alpha$ emission
or the luminosity in the far-ultraviolet or far-infrared (Hippelein et al.\
2003; Engargiola et al.\ 2003; Heyer et al.\ 2004; Gardan et al.\ 2007;
Boissier et al.\ 2007; Verley et al.\ 2009). Kang et al.\ (2012) have
summarised these works and they determine $\psi\sim0.02$ M$_\odot$ yr$^{-1}$
kpc$^{-2}$ within the central square kpc of M\,33, but possibly as much as
0.03--0.04 M$_\odot$ yr$^{-1}$ kpc$^{-2}$ (Heyer et al.\ 2004). This is $\sim10$
times higher than our estimate.}
\end{itemize}
If we shorten the pulsation duration (as used in Paper II) by this factor,
then the fractional mass-loss of stars with masses $M>4$ M$_\odot$
($\log M>0.6$) becomes $\eta\sim0.4$. This reconciles the estimated mass-loss
rates and star formation rates, given that some additional mass will be lost
by the most massive stars as blue supergiants and supernov{\ae}.

%
% FIGURE 17
%
\begin{figure*}
\centerline{\hbox{\hspace{8mm}
\psfig{figure=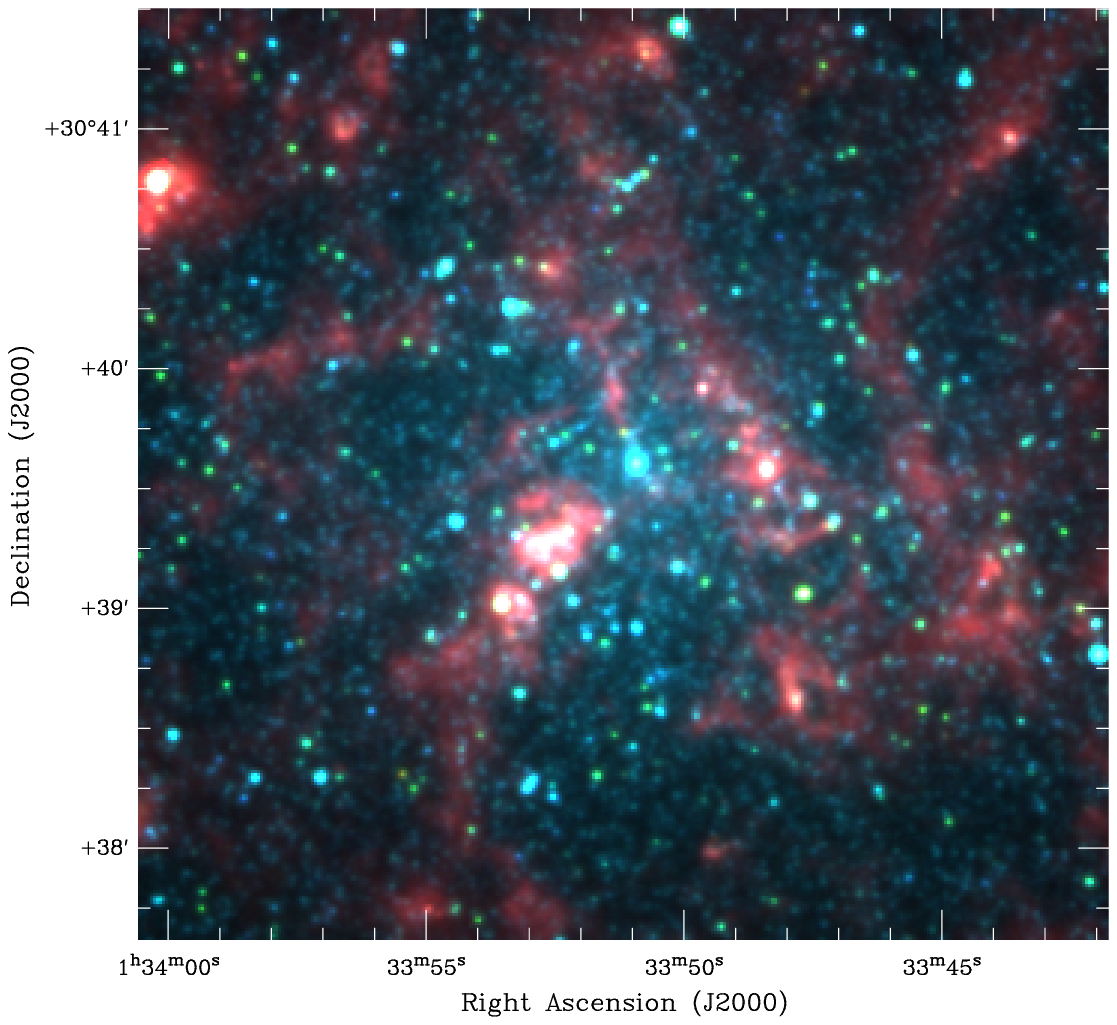,width=74mm}
\psfig{figure=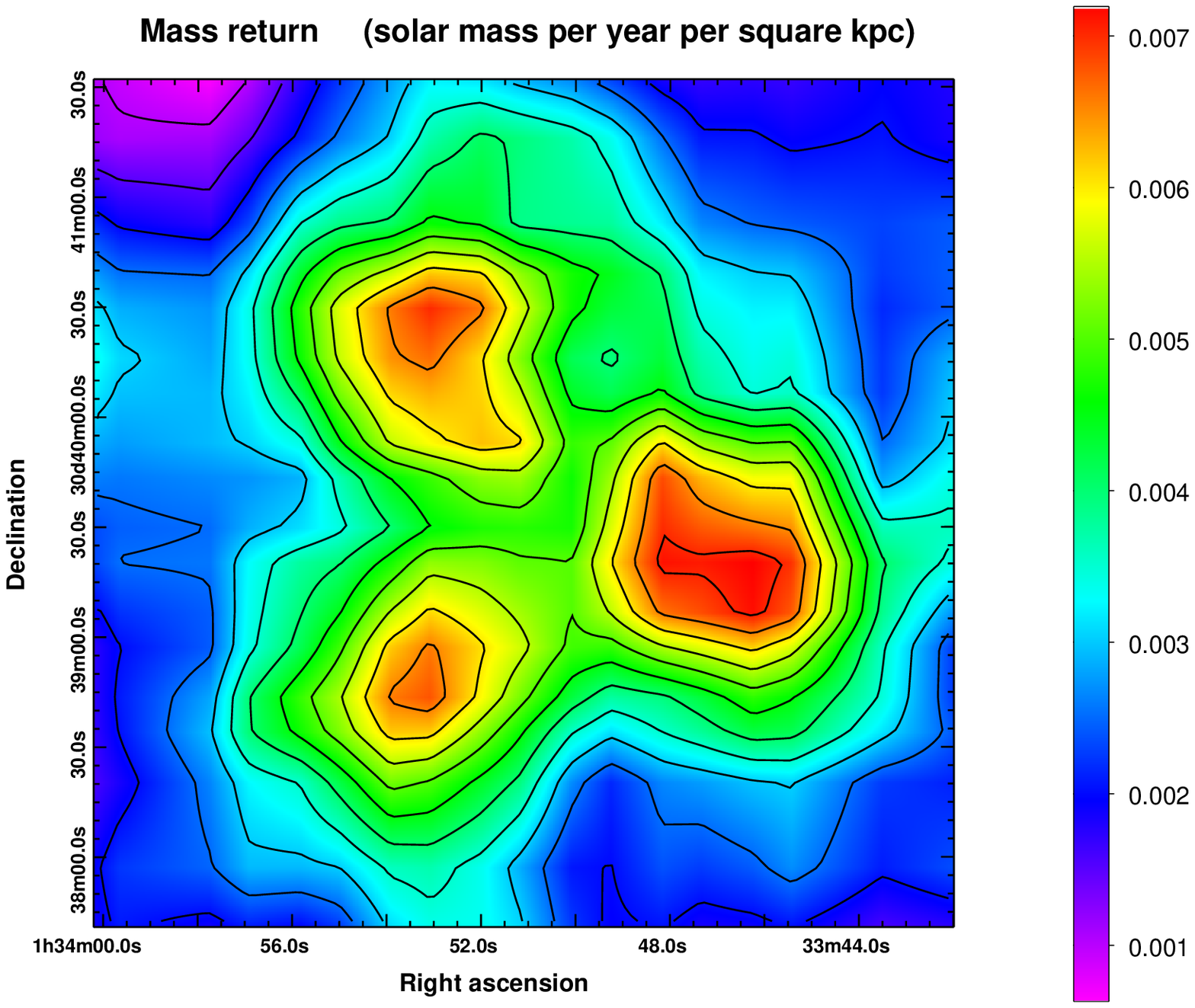,width=94mm}
}}
\caption[]{Left: {\it Spitzer} composite image of IRAC bands 1, 2 and 4 at
respectively 3.6 $\mu$m (blue), 4.5 $\mu$m (green) and 8 $\mu$m (red); Right:
map of mass-return-rate surface density over the central region of M\,33.}
\end{figure*}

Now that we have regained some confidence in the estimated values of the star
formation rate and mass-return rate, we can consider the balance between the
ISM depletion and replenishment. Kang et al.\ (2012) estimated an ISM
depletion timescale of $M_{\rm ISM}/\psi\approx0.3$ Gyr. But this does not take
into account the continuous mass return by evolved stars, which would lengthen
the depletion timescale. Our estimated mass-return rate (after deprojection)
of $\dot{M}_{\rm return}\sim0.004$--0.005 M$_\odot$ yr$^{-1}$ kpc$^{-2}$ (up to
0.01 M$_\odot$ yr$^{-1}$ kpc$^{-2}$ if all carbon stars are actually M-type
stars) is more than our initially estimated recent star formation rate of
$\psi\sim0.003$ M$_\odot$ yr$^{-1}$ kpc$^{-2}$, but lower than the star
formation rate estimated by other groups. We have just argued that our
estimate of the star formation rate needs to be scaled up by a factor
$\sim10$, which would mean that the mass-return rate falls short of supplying
the fuel for continued star formation by a factor 6 or 7 (or 3, if all carbon
stars are actually M-type stars). The naive estimate of the depletion
timescale of 0.3 Gyr would not change by more than 17\% as a result of this
mass return -- and by at most 50\% if all carbon stars turn out to be M-type
stars.

However, additional mass is returned by supernov{\ae}, hot massive-star winds,
luminous blue variable eruptions, et cetera. Most of these contributions come
from massive stars, which contribute about one third to the mass returned
through dusty stellar winds (see Fig.\ 15, top right panel). Their fractional
mass-loss (after correction for the shorter pulsation duration) is $\sim40$\%,
i.e.\ they could at most contribute $\sim2$ times as much as estimated here
(accounting also for stellar remnants being left behind), which would increase
the total mass-return rate (across all stellar masses) from $\sim0.004$--0.005
to $\sim0.006$ M$_\odot$ yr$^{-1}$ kpc$^{-2}$. Note that the contribution from
low-mass stars is already fully accounted for by the dusty stellar winds.
Thus, the above conclusion does not change, namely that the rate at which the
ISM is replenished by stellar mass loss falls short by a factor $\sim5$ to
sustain star formation at the current rate. This conclusion does not change if
all carbon stars turn out to be M-type stars, though the discrepancy is
reduced to a factor $\sim3$. For star formation to continue beyond the next
few hundred Myr gas must flow into the central regions of M\,33, either
through a viscous disc or via cooling flows from the circum-galactic medium.

The mass return is not uniform across the central square kpc, so locally the
above conclusions might not be accurate. The mass return by low-mass stars and
carbon stars is fairly uniform across the area, with a slight radial gradient.
But there are three areas at $r\sim0.3$--0.4 kpc from the centre where the
mass-return rate is much higher due to the contribution of several massive,
very dusty stars. This could feed -- and chemically enrich -- enduring or new
star formation in those areas on timescales of a couple of $10^7$ yr.

%
% FIGURE 18
%
\begin{figure}
\centerline{\vbox{
\psfig{figure=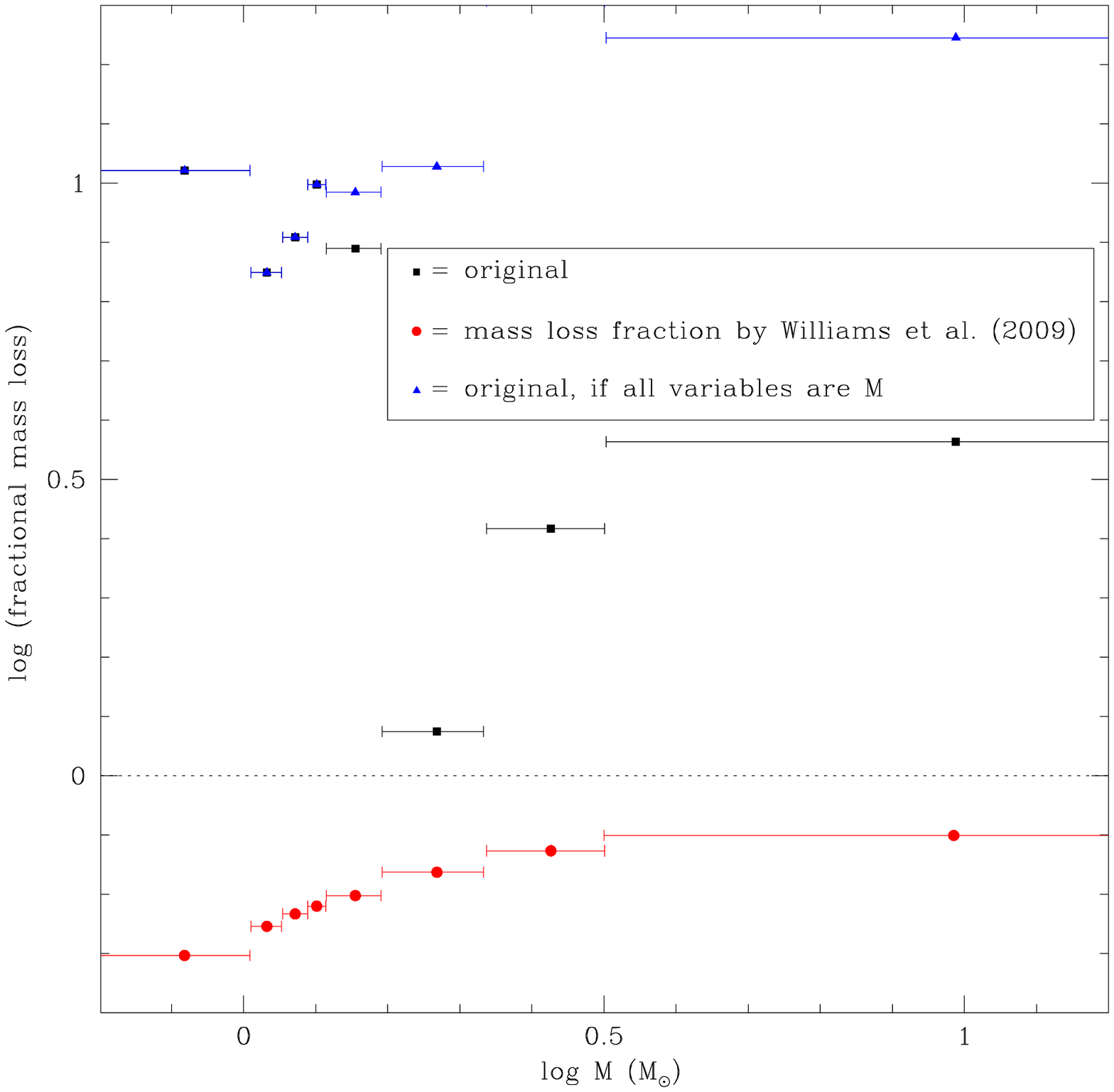,width=84mm}
\psfig{figure=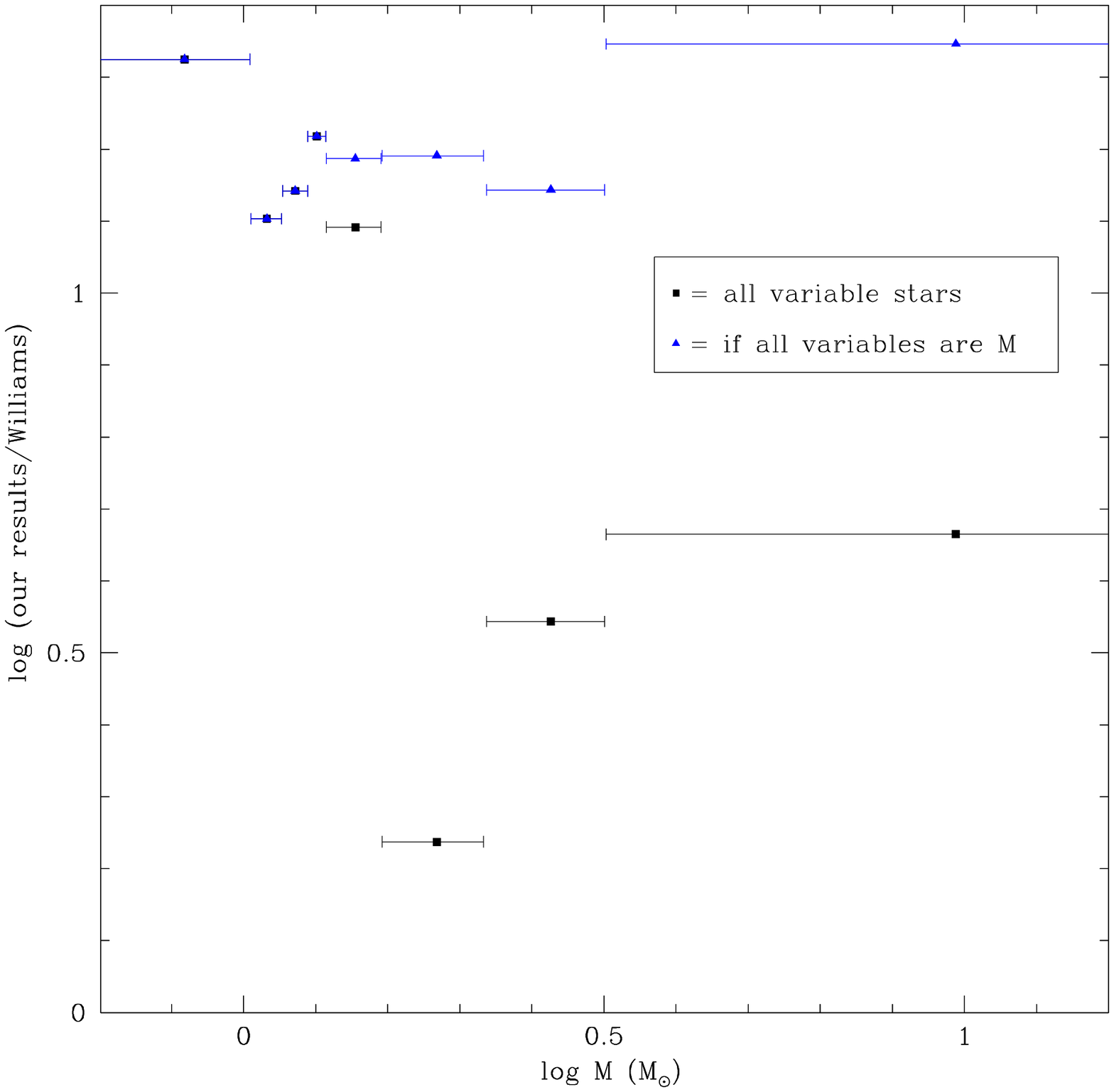,width=84mm}
}}
\caption[]{Top: ratio of the mass lost during the pulsational instability
phase of stellar evolution, to the birth mass ($M$), i.e.\ the fraction of
mass lost in the form of a dusty stellar wind. The dotted horizontal line
marks the fiducial case where the star expels all of its mass. For comparison
we also plot the fractional mass loss derived from the initial--final mass
relation determined by Williams et al.\ (2009). Bottom: ratio of our derived
fractional mass loss to that derived from the initial--final mass relation.}
\end{figure}

In Paper II we had found that the star formation rate peaks within the central
$r<0.2$ kpc. The very centre of M\,33, while characterised by high levels of
mass return is not where the ISM is replenished at the highest rate. However,
if gas from the dense regions at $r\sim0.3$--0.4 kpc can be accreted into the
nucleus of M\,33 then this could be a mechanism for star formation to occur
even within the nuclear star cluster. Indeed, Tosaki et al.\ (2011) find the
molecular gas peaks at $r\sim0.1$ kpc from the centre, and this could have
been supplied at somewhat greater distances before being channeled towards the
nucleus. They find another peak at $r\sim0.6$ kpc which is at the edge of our
survey area. The offsets of the peaks in the dense ISM with respect to the
peaks in mass return (and star formation) suggest that, globally, there is a
delay as well as a migration between the epochs and sites of intense mass
return, accumulation of dense ISM, and star formation.

%=============================================================================
\section*{Acknowledgments}

We thank the staff at UKIRT for their excellent support of this programme. JvL
thanks the School of Astronomy at IPM, Tehran, for their hospitality during
his visits. We are grateful for financial support by The Leverhulme Trust
under grant No.\ RF/4/RFG/2007/0297, and by the Royal Astronomical Society.
Finally, we thank the referee for her/his constructive report which prompted
us to improve the manuscript.

%=============================================================================

\label{lastpage}

\end{document}